# Janus Dipoles: Fundamentals, Realizations, and Emerging Applications

Bo Xue,[1] Qiuyang Li,[1] Yuqiong Cheng,[2] Wenbo Ma,[3] Xuhuinan Chen,[3] Wanting Tong,[4] Junho Jung,[2] Runqi Jia,[1] Ke Chen,[4,a)] Yijun Feng,[4,a)] Xiao Lin,[3,a)] Shubo Wang,[2,a)] and Alex M. H. Wong[1,5,a)]

[1]*Department of Electrical Engineering, City University of Hong Kong, Tat Chee Avenue, Kowloon, Hong Kong 999077, China*

[2]*Department of Physics, City University of Hong Kong, Tat Chee Avenue, Kowloon, Hong Kong 999077, China*

[3]*State Key Laboratory of Extreme Photonics and Instrumentation, Zhejiang Key Laboratory of Intelligent Electromagnetic Control and Advanced Electronic Integration, College of Information Science & Electronic Engineering, Zhejiang University, Hangzhou 310027, China*

[4]*School of Electronic Science and Engineering, Nanjing University, Nanjing, Jiangsu 210023, China*

[5]*State Key Laboratory of Terahertz and Millimeter Waves, City University of Hong Kong, Tat Chee Avenue, Kowloon, Hong Kong 999077, China*

[a)]*Authors to whom correspondence should be addressed: ke.chen@nju.edu.cn; yjfeng@nju.edu.cn; xiaolinzju@zju.edu.cn; shubwang@cityu.edu.hk; and alex.mh.wong@cityu.edu.hk*

## Abstract

The Janus dipole – featuring orthogonally oriented electric and magnetic dipoles with a 90-degree phase difference – has emerged as a powerful paradigm for wave manipulation. Unlike traditional Huygens dipoles used for directional control, this unique configuration exhibits strongly asymmetric, face-selective near-field behavior while maintaining a quasi-isotropic far-field radiation pattern. These remarkable properties make the Janus dipole an essential platform for directional wave shaping, with wide-ranging applications in on-chip photonics, quantum interactions, and wireless power transfer. This review systematically traces the rapid development of the Janus dipole from its foundational theoretical inception to its diverse implementation platforms across optical, microwave, and acoustic frequencies. In this paper, we explore the governing principles, classify realization strategies into passive Janus dipoles, active Janus dipoles, and advanced near-field coupling control, and highlight emerging frontiers. By bridging foundational electrodynamics with advanced device engineering, this paper serves as an essential reference and roadmap for researchers designing next-generation, highly integrated, and compact wave-manipulation systems.

## 1. Introduction

Controlling the direction of wave propagation – including electromagnetic, acoustic and other coherent waves – remains a cornerstone of modern physics and engineering, underpinning applications ranging from on-chip photonic routing,[1–6] quantum light-matter interactions[7–9] to wireless power transfer[10–12] and sensing[13–16]. Achieving selective and tunable directionality is challenging yet essential for future integrated and multifunctional devices. In the near-field regime, directionality refers to the preferential coupling of evanescent energy toward a specific side, which is quantitatively characterized by the ratio of the coupled energy to the non-coupled energy.

Directional dipoles have emerged as the fundamental building blocks for such control. Three directional dipoles have been proposed: the circular dipole, which exploits spin-momentum locking for helicity-dependent unidirectional near-field coupling;[7,17–24] the Huygens dipole, which produces unidirectional near-field coupling and far-field power flow through in-phase electric-magnetic interference;[25–35] and the Janus dipole, whose near-field directionality originates from the imaginary part of the Poynting vector (reactive power).[36–55] Among these, the Janus dipole stands out for its distinctive construction and properties: it consists of orthogonally oriented electric and magnetic dipoles with a 90° phase difference, resulting in strongly asymmetric, face-selective near-field behavior. This enhances coupling to one side while suppressing the coupling to the opposite side of the source. Meanwhile, a quasi-isotropic far-field radiation characteristic is maintained. This unique combination of strongly directional near-field behavior and quasi-isotropic far-field emission enables versatile near-field directional control and opens up new avenues for future integrated applications.

Since its theoretical proposal in 2018,[36] the Janus dipole has undergone rapid development, progressing from a purely conceptual dipolar model to diverse implementation platforms in optics, microwaves and acoustics. Early realizations primarily relied on passive mechanisms, while subsequent advances introduced active configurations and near-field manipulation approaches to improve efficiency, bandwidth, and tunability. As summarized in Fig. 1, these developments can be grouped into three categories: passive Janus dipoles,[42–44] active Janus dipoles,[45–48] and near-field coupling control (via polarization modulation and reconfigurable waveguides).[49–54] Passive Janus dipoles are induced by external illumination of resonant scatterers, where orthogonal electric and magnetic responses emerge under carefully chosen excitation conditions (wavelength, polarization, and/or incident angle). This approach offers straightforward proof-of-concept demonstrations but is often limited by narrow bandwidth, low coupling efficiency, and sensitivity to excitation parameters. Active Janus dipoles incorporate driven sources to generate intrinsic Janus behavior, independent of external excitations. This category achieves significantly higher power efficiency, broader bandwidth, and greater control flexibility, making it particularly suitable for practical device integration. Furthermore, controlling the near-field coupling of Janus dipoles – such as through polarization modulation or reconfigurable waveguides – unlocks unprecedented flexibility, extending the concept to dynamic and programmable scenarios. These three categories reflect the evolution of Janus dipole research toward more efficient, versatile, and application-oriented directional platforms.

Building on the rapid progress in realization methods, the Janus dipole offers compelling application value through its ability to deliver strong selective near-field directionality while preserving quasi-isotropic far-field patterns. In compact Multiple-Input Multiple-Output (MIMO) antenna systems, for instance, the Janus dipole's asymmetric near-field coupling can suppress mutual coupling between adjacent elements by orders of magnitude, allowing dense array packing without degrading the quasi-isotropic radiation pattern required for omnidirectional coverage and MIMO diversity. This dual characteristic further supports selective near-field routing in integrated photonic circuits (e.g., one-sided waveguide excitation with minimal crosstalk), directional emission in quantum optics, near-field energy transfer, and controlled wave manipulation in acoustic and optical trapping systems. Integration with reconfigurable metasurfaces also promises dynamic switching of directional radiation or propagation. Collectively, these potentials position the Janus dipole as a promising candidate for next-generation compact, high-performance, and multifunctional wave devices.

This review provides a comprehensive overview of the Janus dipole, from its theoretical foundations to various realizations and emerging trends. It systematically covers the underlying physics (from parity considerations and an electromagnetic wave perspective), realization strategies across the three main categories (passive, active, and near-field coupling control), key experimental demonstrations, and connections to related concepts such as moving directional sources, higher-order Janus dipoles, quasi-isotropic antennas, Janus trapping and Janus metasurfaces. By synthesizing contributions from collaborative groups and highlighting recent advances, this review represents a unified reference for researchers in physics, nanophotonics, antenna engineering and acoustics, while outlining promising directions for future development. Looking forward, we envision that the Janus dipole's unique combination of strong near-field directionality and desirable far-field properties will establish it as a key enabling technology for next-generation integrated, compact, and multifunctional systems in nanophotonics, antenna engineering, and related fields.

## 2. Fundamental Principles of Janus Dipoles

The theoretical foundation of the Janus dipole lies in the interplay between its orthogonal electric and magnetic dipole moments with a 90° phase difference. This unique configuration gives rise to distinctive near-field and far-field characteristics that cannot be fully captured by conventional Huygens or circular dipoles. To understand its near-field directionality mechanisms, three complementary perspectives are particularly insightful: the polarization matching perspective, the symmetry matching view, and the electromagnetic wave perspective based on the Poynting vector.

### 2.1 Polarization Matching Perspective

From the viewpoint of polarization matching, the out-of-phase orthogonal electric and magnetic fields of an evanescent wave give rise to the rotation of hybrid electric-magnetic polarization,[56] and the associated reactive power quantity resembles a synthetic spin angular momentum.[36] Analogously, the out-of-phase orthogonal electric and magnetic components in a Janus dipole define its dipole polarization state. Once the spin of the Janus dipole matches or opposes that of the evanescent wave, the Janus dipole can exhibit asymmetric coupling to the guided or surface modes. This process can be quantitatively described by Fermi's golden rule:[2,19,57,58]

$$\kappa_{\pm} = \left|\mathbf{p}^{*} \cdot \mathbf{E}_{\pm} + \mathbf{m}^{*} \cdot \mu \mathbf{H}_{\pm}\right|^{2} \tag{1}$$

where $\mathbf{E}_{\pm}$ and $\mathbf{H}_{\pm}$ are electric and magnetic fields of the guided or surface modes at the location of the Janus dipole, $\mu$ is the permeability of the surrounding medium, and the subscript $\pm$ denotes the opposite evanescent decay directions of the modes, i.e., the directions of the reactive power vector. For planar guided or surface modes supporting evanescent plane waves, Eq. (1) is analytically equivalent to the angular-spectrum description of dipole scattering in momentum space[17,22] and provides direct physical insight into the near-field dipolar interference underlying the coupling asymmetry. For structured guided modes, however, the evanescent tails possess a more complex texture of vector fields[59] and Eq. (1) generally requires numerical evaluation. The coupling coefficients $\kappa_+$ and $\kappa_-$ respectively quantify the polarization overlap between the Janus dipole and two oppositely decaying evanescent waves, and they directly determine the near-field directionality, which can be defined as

$$D = \frac{\kappa_+ - \kappa_-}{\kappa_+ + \kappa_-} \tag{2}$$

Importantly, when the Janus dipole is perfectly mismatched to one evanescent wave, unidirectional coupling can be achieved, i.e., coupling vanished on one side ($\kappa_+ = 0$) while remaining on the other side ($\kappa_- \neq 0$), resulting in a directionality of unity magnitude.[36,60] This polarization-based coupling framework enables a straightforward route for designing directional Janus dipoles for arbitrary waveguide configurations. It can also be extended to higher-order multipolar sources, achieving more versatile control of near-field mode excitation.[40,41,61]

### 2.2 Symmetry Matching Perspective

From the viewpoint of symmetry matching, the near-field coupling between a Janus dipole and guide or surface modes is also governed by their symmetry properties.[36,45] As summarized in Table 1, an electric dipole, represented as a pair of opposite electric charges separated by a distance, is odd under parity ($\hat{P}$) symmetry transformation and even under time-reversal ($\hat{T}$) symmetry transformation. In contrast, a magnetic dipole, visualized as a current loop, is even under $\hat{P}$ transformation and odd under $\hat{T}$ transformation. Consequently, a Janus dipole formed by out-of-phase electric and magnetic dipoles is odd under $\hat{P}$ transformation and even under $\hat{T}$ transformation. Similarly, for an evanescent wave, the electric field is odd (even) under $\hat{P}$ ($\hat{T}$) transformation, whereas the magnetic field is even (odd) under $\hat{P}$ ($\hat{T}$) transformation. Therefore, the reactive power vector associated with the out-of-phase electric and magnetic fields is also odd under $\hat{P}$ transformation and even under $\hat{T}$ transformation. The symmetry correspondence between the Janus dipole and the reactive power vector of the evanescent wave reveals that Janus-type directionality fundamentally depends on the evanescent decay direction. Since a pair of counterpropagating modes in one waveguide carry the same reactive power due to $\hat{T}$ symmetry, the side of the Janus dipole facing the waveguide determines whether the coupling to the waveguide is completely suppressed or allowed, referred to as face-dependent directional coupling.[36,45] This symmetry-based framework generalizes the physical picture of the Janus source beyond a specific electric-magnetic dipole realization and identifies internal symmetry as a key factor for directional manipulation. In particular, Ref.[45] demonstrated all-electric subwavelength metasources for Janus-type light routing without introducing magnetic dipoles and realized symmetry-selective near-field directionality, which would broaden practical scenarios of Janus source from information communications to quantum interactions.

**2.3 Electromagnetic Field Perspective**

From the electromagnetic field perspective, the near-field directionality of the Janus source can be understood by analyzing the imaginary part of its Poynting vector. For a Janus dipole as shown in Fig. 2(a), which is composed of a z-directed electric Hertzian dipole and an x-directed magnetic Hertzian dipole with a 90° phase difference, the electromagnetic field and Poynting vector can be analytically derived. Refs.[47,48] detail the derivation of individual electromagnetic near-field components. Specifically, the near-field coupling characteristics can be understood by examining the complex Poynting vector $\mathbf{S} = \frac{1}{2}\mathbf{E} \times \mathbf{H}^* = \mathrm{Re}\{\mathbf{S}\} + j\mathrm{Im}\{\mathbf{S}\}$ in the near-field regime. Generally, the real part $\mathrm{Re}\{\mathbf{S}\}$ represents the time-averaged power radiated to the far field, which remains quasi-isotropic and exhibits no significant directionality. Fig. 2(b) presents the calculated far-field radiation pattern of the Janus source in the xoy-plane. Radiation to the z-direction is the same as the x-direction. In contrast, the imaginary part $\mathrm{Im}\{\mathbf{S}\}$, which quantifies the reactive power density, predominantly determines the near-field directionality of the Janus dipole. Near the source ($kr \ll 1$), the reactive power $\mathrm{Im}\{\mathbf{S}\}$ dominates and yields a complex, asymmetric spatial profile:

$$\mathrm{Im}\{\mathbf{S}\} \propto \left(\frac{2\sin\phi}{(kr)^2} - \frac{\cos^2\phi}{(kr)^3} - \frac{\sin\phi}{(kr)^4}\right)\hat{\mathbf{r}} + \left(\frac{\cos\phi}{(kr)^4} - \sin\phi\cos\phi\left[\frac{1}{kr} + \frac{1}{(kr)^3}\right]\right)\hat{\boldsymbol{\phi}} \tag{3}$$

where $k$ represents the wavenumber of the host medium, $r$ and $\phi$ denote the radial distance and azimuthal angle in standard spherical coordinates with respect to the source, and $\hat{\mathbf{r}}$and $\hat{\boldsymbol{\phi}}$ represent the corresponding unit vectors.

We make a few observations regarding the $\mathrm{Im}\{\mathbf{S}\}$, which represents the reactive power density associated with the electromagnetic near-field. We can see that there are three terms along the r-direction – one varying with $\cos^2\phi$, and the other two related to $\sin\phi$. The $\cos^2\phi$ term dominating in the extreme near-field is symmetric in both x- and y- directions. However, the remaining two terms which vary with $\sin\phi$ are noteworthy for their asymmetry in the y-direction. Fig. 2(c) plots the imaginary Poynting vector for the Janus source, showing that the vector field has a directed flow along the +y axis with high intensity concentrated near the source location. The direction of this coupling is dictated by the orientation of $\mathrm{Im}\{\mathbf{S}_J\}$, which in turn depends on the 90° phase lead or lag between the electric and magnetic dipoles: a phase lead in the electric current relative to the magnetic current directs $\mathrm{Im}\{\mathbf{S}\}$ along +y, while a lag reverses it to -y. We note that the direction of $\mathrm{Im}\{\mathbf{S}\}$ is intrinsically linked to reactive (inductive or capacitive) nature of the source. Since $\angle\mathbf{S} = \angle\frac{1}{2}\{\mathbf{E} \times \mathbf{H}^*\} = \angle\mathbf{E} - \angle\mathbf{H}$, when $\angle\mathbf{E} - \angle\mathbf{H} = 90°$, $\angle\mathbf{S} = 90°$ and $\mathrm{Im}\{\mathbf{S}\}$ is maximized in the +y-direction. In this case, the near-fields of the source is inductive, as $\eta = \frac{E}{H}$ has a phase of 90°. Conversely, when $\angle\mathbf{E} - \angle\mathbf{H} = -90°$, $\mathrm{Im}\{\mathbf{S}\}$ is maximized in the -y-direction, and in this case, the near-fields of the source is capacitive, as $\eta = \frac{E}{H}$ has a phase of $-90°$. This aspect of the Poynting vector's direction characteristic strongly affects the near-field coupling dynamics of the Janus source, and can be seen as the cause of near-field coupling effects of Janus sources with dielectric waveguides observed in earlier works.[36,38,42–51,54]

## 3. Realization and Near-Field Manipulation of Janus Dipoles

Following the seminal demonstration of the Janus dipole concept,[36] numerous studies[42–54] have explored its realization through both numerical simulations and experimental platforms to validate and exploit its distinctive near-field directionality. Depending on the design strategies utilized to generate the required orthogonal electric and magnetic dipole moments (with 90-degree phase difference) and tuning mechanisms, we classify these implementations into three primary categories: passive Janus designs,[42–44] active Janus configurations,[45–48] and advanced near-field manipulation (through polarization modulation[49–51] and reconfigurable waveguides[52–54]). In the following subsections, we review representative implementations in each category, the underlying dipole construction approaches, and the experimental and/or simulated evidence for their near-field directionality.

### 3.1 Passive Janus dipoles

The initial works on Janus sources investigated idealized electric and magnetic dipoles which could not be directly validated experimentally. To address this, subsequent studies[42–44] generate Janus sources by illuminating carefully designed passive structures. Under external illumination (optical or electromagnetic plane waves), these particles induce electric and magnetic dipole moments. By tuning the incident angle, polarization and/or frequency of the external illumination, the directions, amplitudes, and phase differences between the induced electric and magnetic dipole moments can be controlled, enabling the construction of the Janus source. We term this class of structures the *passive* Janus sources to emphasize their reliance on external excitation to emulate Janus behavior, distinguishing them from intrinsic theoretical constructs and the *active* Janus sources which we review at a later section.

Ref.[43] elaborates and analyzes the near-field directionality of Janus sources under the acoustic scattering mechanism. In photonics, Janus sources utilize the special amplitude and phase relationship between the generated electric and magnetic fields to achieve unidirectional waveguide coupling. In acoustics, there is similar amplitude and phase relationship between the pressure and the velocity fields. Janus sources are realized by using a combination of acoustic monopole $M$ and dipole $\boldsymbol{D}$. Their acoustic pressure field angular spectrum is: $p(\boldsymbol{k}) = \frac{i}{2\pi k_0 k_z}(M + \frac{\boldsymbol{k}}{k_0} \cdot \boldsymbol{D})$. Considering their coupling with a planar waveguide in the z-direction, to achieve the acoustic Janus source, the relationship between the monopole and dipole is $M = \pm i\gamma D_Z = \pm i\sqrt{n_{eff}^2 - 1}\, D_Z$, (where $n_{eff}$ represents the effective refractive index of the waveguide, and the positive and negative signs represent the source coupling direction). For example, when $M = +i\gamma\, D_Z$, in the +z direction, $p \propto M + \frac{k_z}{k_0} D_Z = i\gamma D_Z + \frac{i\gamma k_0}{k_0} D_Z = 2i\gamma D_Z \neq 0$, and there is coupling on this side; in the negative z direction, $p \propto M + \frac{k_z}{k_0} D_Z = i\gamma D_Z - \frac{i\gamma k_0}{k_0} D_Z = 0$, and there is complete cancellation on this side. To obtain an acoustic Janus source, a high acoustic index cylinder was designed, as shown in Fig. 3(a). It can simultaneously generate acoustic monopoles and dipoles under plane acoustic wave illumination. The relative intensity and phase between the monopoles and dipoles can be controlled by manipulating the cylinder's material parameters geometric dimensions. Simulation results in Figs. 3(b) demonstrates such a cylindrical Janus source. When the waveguide is above it, the waveguide is excited; when it is below, almost no excitation occurs. This verifies the near-field directional coupling of the acoustic Janus source.

Ref.[42] addresses the limitations of previous nanoresonators, which relied on strict excitation conditions (e.g., plane waves at specific angles and wavelengths). To overcome this, the authors propose a theoretical and computational framework based on quasinormal modes (QNMs). QNMs are the natural resonances of an object, possessing an intrinsic (excitation-independent) multipole moment. After an analysis of the QNMs, the resonator shape can be designed to selectively excite the QNM with specific electric and magnetic dipole moments to synthesize the Janus source. Following this method, a dolmen nanoresonator is designed, as depicted in Fig. 3(c). Fig. 3(d) shows the near-field coupling between the dolmen scatterer and the nanowire before and after the dolmen scatterer rotating 180°. The scattered light can be selectively coupled to one side of the waveguide, exhibiting the near-field directionality of the Janus source. Furthermore, this coupling method is robust over a wide angle; the near-field directivity remains stable even when the incident angle of light varies over a large range (up to about 80°).

Ref.[44] proposes the directional dipole dice (DDD) – a unified platform capable of synthesizing all three dipoles (the circular, Huygens and Janus dipoles) and constructing a complete directional space. Fig. 3(e) illustrates the proposed DDD. It is a metallic helical structure that, under tailored plane-wave illumination, can selectively excite all three fundamental dipole modes. Crucially, each dipole's near-field directionality can be reversed via incident angle modulation. Fig. 3(f) displays the electric field ($E_x$) distributions in the waveguide when the helical structure is positioned below and above the waveguide, respectively. As clearly shown, guided waves are only effectively excited when the optimized Janus dipole is placed beneath the waveguide. Authors experimentally demonstrated the directional coupling characteristics of the DDD in the microwave regime. A copper helix is designed with dipole resonances at 2.35 GHz. Plane-wave excitation is delivered by a horn antenna. The plane-wave excitation induces corresponding dipole moments in the helix, which then couple waves directionally into the dielectric waveguide. The experimental results show that the near-field directivity of the Janus dipole reaches a maximum value of 32, which means that about 97% of the power is coupled to the upper waveguide.

### 3.2 Active Janus dipoles

Though passive Janus dipoles are conceptually viable, in most cases the particles only re-scatter a very small part of the incident wave, leading to low power efficiency. Further, the re-scattering amplitude and phase of the electric and magnetic dipoles are carefully optimized for a single frequency, leading to single-frequency and/or narrowband directionality phenomena. To overcome these limitations, subsequent efforts have explored active Janus sources, which directly generate the required orthogonal electric and magnetic dipoles with precisely controlled amplitude, phase, and orientation, independent of external illumination. Compared to the passive counterparts, active Janus sources promise significantly higher efficiency and more flexible modulation capabilities than their passive counterparts.

The first investigation into the active Janus effect within the acoustic domain is presented in Ref.[46]. In this work, the researchers successfully construct an acoustic Janus source at a frequency of 2 kHz and observe its near-field coupling directionality. The acoustic Janus source is constructed by combining the complex excitation for 5 speakers as shown in Fig. 4(a). The phase and amplitude are controlled by external circuitry. Similar to the optical system, in order to observe the near-field distribution characteristic of the source, a special acoustic metamaterial (with effective negative mass density and effective phase velocity) is used as media and placed above and below the source speakers. Figs. 4(b) illustrates the near-field directional characteristics of the Janus source. Specifically, by applying appropriate excitations to the speakers $\phi_0$, $\phi_2$, and $\phi_4$ (as indicated in the schematic of Fig. 4(a)), the Janus source can be successfully excited, resulting in asymmetric near-field coupling. Furthermore, the directionality can be configured by adjusting the phases and amplitudes of $\phi_2$ and $\phi_4$, demonstrating the flexibility and control offered by the speaker array configuration in realizing the Janus source behavior.

Active Janus sources have been explored in the microwave domain at 2.7 GHz.[45] As shown in Fig. 4(c), this metasource consists of five subwavelength electric sources, which, when appropriately excited, form an active Janus dipole. This Janus dipole couples with a transmission line metamaterial that is composed of epsilon-negative (ENG) and mu-negative (MNG) unit cells, exhibiting near-field directionality of the excited guided surface wave. Fig. 4(d) presents the coupling results from both simulations and experiments, revealing selective excitation only to one of the two sides. This work lays the groundwork for more versatile and integrated implementations of Janus sources in higher-dimensional or free-space scenarios.

Despite the promising demonstrations of active Janus sources, these early implementations rely on multi-element arrays and have not yet provided concrete application scenarios for practical electromagnetic systems. To bridge this gap, a practical active Janus source based on the twin current filament (TCF) model embedded within a parallel-plate waveguide (PPW) is demonstrated, as shown in Fig. 5(a).[47] The TCF configuration consists of two closely spaced current filaments, which generate co-located orthogonal electric and magnetic currents with controlled amplitude and 90° phase difference through precise tuning of the filament currents. This source can be excited using two monopole antennas fed by a simple microwave circuit. For near-field characterization, two dielectric waveguides are placed adjacent to the TCF source on opposite sides. Simulations confirm the presence of a strong near-field directionality, selectively coupling to the waveguide on one side of the source (e.g., the +y direction) while effectively suppressing coupling to the opposite side (e.g., the -y direction) (Fig. 5(b)). Furthermore, the coupling direction can be readily reversed by adjusting the orientation of the magnetic dipole, highlighting the flexibility of the TCF configuration. A key potential application is in compact MIMO antenna systems, where closely spaced elements commonly suffer from severe mutual coupling, degrading overall performance. Conventional approaches, such as decoupling networks[62–64] or pattern diversity methods,[65–67] achieve varying degrees of success but may introduce feed network complexity and angular variations in communication quality. By leveraging the Janus source's asymmetric near-field coupling and quasi-isotropic radiation, this work demonstrates that two active Janus sources placed in close proximity (0.08 to 0.24 wavelengths) exhibit mutual coupling suppression of approximately 1000-fold compared to conventional dipole pairs at similar separations. The model setup for two such sources in the PPW is illustrated in Fig. 5(c). Notably, even at the very small spacing of $d = 0.08\ \lambda_0$, the mutual coupling between the two active Janus sources remains weak. Simulations show low coupling strength in a color-coded plot comparing the Janus source with Huygens and electric dipole pairs (right panel of Fig. 5(d)). Experiments further confirm this isolation via S-parameters, where $S_{21}$ (representing the coupling coefficient between the two sources) is significantly suppressed compared to conventional dipoles (left panel of Fig. 5(d)). This substantial isolation is realized without compromising the radiation properties of individual antennas, highlighting the potential for power-efficient active directional near-field devices and a new approach to compact MIMO antenna design.

While the two-dimensional active Janus sources discussed earlier demonstrate effective near-field directionality within confined waveguide environments, they remain limited to planar implementations and single-frequency, constraining their extension to free-space electromagnetic applications. To overcome these limitations, a three-dimensional active Janus source – referred to as a Janus antenna – is realized, representing a class of quasi-isotropic antennas that exhibit strong near-field directionality in addition to their well-known far-field quasi-isotropic radiation patterns.[48] This configuration is constructed by superposing orthogonal electric and magnetic dipoles with a 90° phase difference, as illustrated in Fig. 5(e). The antenna incorporates capacitively loaded loops (CLLs), a folded dipole, and coplanar stripline section (CPS) on a dielectric substrate to achieve the required dipole moments. Unlike prior works focused solely on far-field radiation characteristics of quasi-isotropic antennas, this study reveals their previously unexplored Janus-like near-field properties. As shown in Fig. 5(e), near-field directionality is demonstrated by exciting the antenna (Port 1) and observing coupled electric field magnitude in two dielectric waveguides (Ports 2/4 on one side, Ports 3/5 on the other). The optimized Janus antenna directs most (98%) of the coupled power to the preferred (-x) side, achieving a directionality difference approaching 20 dB between the coupled and uncoupled ports as shown in Fig. 5(f). A 3-dB or stronger directionality is maintained over a 28.9% bandwidth (2.28 to 3.04

GHz) for the optimized Janus antenna. Remarkably, the power efficiency – defined here as the ratio of coupled power to the target waveguide port under identical input excitation power – is enhanced more than 105-fold compared to exciting a passive Janus dipole with a horn antenna in a similar setup, as detailed in[44]. This enables comparable waveguide coupling with significantly lower input power. This work bridges the physics community's investigations of Janus dipoles (emphasizing near-field reactive power asymmetry) and the antenna community's studies on quasi-isotropic radiators (emphasizing far-field isotropy), culminating in a wideband, power-efficient 3D active Janus source with strong potential for compact directional systems in free space.

### 3.3 Controlling near-field coupling of Janus dipoles

While the passive and active Janus designs reviewed in the previous sections successfully establish near-field directionality through specific structural configurations or source setups, recent advancements demand more advanced, real-time control over these coupling behaviors.[9,45,68–70] Controlling the near-field coupling of Janus dipoles unlocks unprecedented flexibility in steering light at the nanoscale, entirely avoiding post-fabrication of the built-in optical sources or rigid nanostructures.[71–73] As illustrated in Fig. 6 and Fig. 7, there are various dynamic near-field coupling mechanisms of dipolar sources,[49,50,54,74] including controlling near-field coupling of Janus dipole via polarization modulation[49,54] in Fig. 6 and reconfigurable waveguides[52] in Fig. 7.

#### 3.3.1 Controlling near-field coupling of Janus dipoles via polarization modulation

Polarization, an inherent property of light, significantly enriches the physics of light-matter interactions. It serves as a powerful paradigm for tailoring optical fields at the nanoscale, enabling the precise manipulation of both the directionality and strength of near-field coupling.[16,75–77] Polarization-controlled near-field interactions are foundational to a wide array of intriguing optical phenomena and advanced optical applications, spanning singular optics,[78–80] near-field microscopy,[57,81] chiral quantum optics,[9,24,82–84] photonic computation,[85] and nanorouters.[86]

Controlling the polarization state of the incident wave offers a direct approach to dictate the near-field coupling of dipolar sources. For instance, spin-momentum locking enables the dynamic flipping of surface-wave excitation directionality by toggling the incident polarization between left-handed and right-handed states.[24,87–91] Alternatively, this directional control can be achieved by using chiral particles of opposite chirality under linearly-polarized excitation.[44,88,92–95] For the Janus dipole, polarization modulation unlocks profound control over its characteristic near-field coupling and non-coupling faces.[49,50,54] As experimentally demonstrated in Figs. 6(a, b), by tailoring the incident illumination (e.g., using linearly-polarized or circularly-polarized light), a nanostructure can be precisely excited as a linear or spinning Janus dipole.[49] Placing the nanostructure on a high refractive index substrate, the TM (i.e., transverse-magnetic or p-polarized) and TE (i.e., transverse-electric or s-polarized) evanescent waves carried by Janus dipoles can be measured through their angular spectra. Consequent to polarization locking, the TM evanescent waves spectrum of these Janus sources characterizes the coupling faces, while the TE evanescent waves spectrum features the non-coupling faces. This distinct behavior establishes a robust mechanism for polarization-selective near-field coupling and the efficient spatial sorting of optical modes.[49]

A representative implementation of this concept was further demonstrated in an integrated silicon-photonic platform, where a silicon nanocylinder was placed between two identical single-mode silicon waveguides to act as a compact Janus dipolar source, as shown in Fig. 6(c).[50] Under TE-polarized illumination near the wavelength of 1.55 μm, the nanocylinder supports orthogonal electric and magnetic dipoles with the required amplitude and phase relation, thereby selectively coupling light into one waveguide while strongly suppressing coupling to the other, as shown in Fig. 6(d). By contrast, TM-polarized illumination does not satisfy the Janus condition and produces nearly balanced excitation of both waveguides. The coupling selectivity can therefore be controlled by the incident polarization, and further tuned by the wavelength and angle of incidence. This work establishes incident-wave polarization as a practical external knob for selective on-chip routing based on Janus dipolar near-field coupling.

Building upon incident wave modulation, near-field interactions can be further tailored by directly manipulating the constituent dipole moments of the sources. By designing these dipole moments, the single-polarization perfect Janus dipole can be achieved, where the excited surfaces wave of a specific polarization is completely suppressed at the non-coupling face.[44,49] Furthermore, by judiciously co-designing the constituent dipole moments and the environmental waveguide dispersion, both the TM and TE surface waves can be simultaneously suppressed at the non-coupling faces, realizing a dual-polarization perfect Janus dipole.[55]

Despite these advances, the comprehensive manipulation of near-field coupling via environmental polarization remains largely underexplored, primarily because most existing research has focused on linearly-polarized systems, where TM and TE evanescent waves are exploited independently. A promising alternative is offered by hybrid TM-TE surface waves, which are supported by symmetry-broken physical systems such as anisotropic metasurfaces,[96–98] chiral platforms (e.g., moiré graphene),[99–101] Dyakonov surface wave interfaces,[102,103] and structures supporting hyperbolic shear polaritons.[104] Unlike linearly-polarized surface waves, hybrid modes can simultaneously extract both the TM and TE evanescent components from a dipolar source.[51] By harnessing these orthogonal evanescent components, hybrid surface waves enable near-field coupling beyond traditional polarization locking. For example, by placing a Janus dipole in a parallel-plate waveguide supporting hybrid modes, its coupling and non-coupling faces can be flipped simply by rotating the dipole within a plane parallel to the interface, as shown in Figs. 6(e, f).[51] This approach unlocks the potential of hybrid TM-TE surface waves and offers a highly versatile alternative scheme for shaping near-field interference of Janus dipoles.

**3.3.2 Controlling near-field coupling of Janus dipoles via reconfigurable waveguides**

Building upon the polarization-controlled near-field interactions discussed previously, dynamic modulation of near-field coupling can also be achieved by actively tuning the properties of the surface waves supported by the local environment. Due to polarization locking, when a Janus dipole's coupling face is oriented upwards for TM waves, its coupling face for TE waves points downwards. Therefore, by placing a Janus dipole between waveguides capable of dynamically supporting either TE or TM surface waves, the near-field coupling can be actively modulated. This active modulation achieved by dynamically tuning a system's bias is highly preferable to physically modifying solid nanostructures, as it avoids crosstalk and minimizes material dissipation.[52,54,105] Reconfigurable environments, such as graphene-metasurface waveguides shown in Fig. 7(a), provide a versatile platform for dynamical modulation through polarization locking.[106–110] By tuning the graphene's chemical potential (i.e., $\mu_c$) via electrostatic gating, the system can dynamically switch the polarization state of the supported surface waves. This environmental reconfiguration fundamentally alters the near-field coupling of various complex dipolar sources. For example, a specifically constructed bi-state dipole acts as a standard circularly-polarized electric dipole when the environment supports TM surface waves. However, when the chemical potential is adjusted to support TE surface waves, that exact same physical source counterintuitively transforms into a Huygens dipole.[54] For a Janus dipole, tuning the waveguide to support TM surface waves actively directs the coupling face upward. Conversely, switching the waveguide to support TE surface waves dynamically flips the near-field coupling face downward,[54] as shown in Fig. 7(b). Thus, active modulation of near-field directionality is achieved entirely without altering the excitation source.

Beyond the polarization state, the group velocity of the supported surface waves offers another powerful degree of freedom to govern the near-field directionality of dipolar sources.[52,53] Because near-field directional routing is fundamentally tied to the energy flow determined by the direction of the group velocity, altering the sign of the group velocity relative to the phase velocity can completely reverse the near-field directionality of the dipolar sources. This principle is elegantly realized in reconfigurable environments like multilayer graphene-hBN heterostructures in Fig. 7(c). This transition dynamically changes whether the excited polaritons exhibit a negative or positive group velocity relative to their phase velocity. For example, placing a Huygens dipole on the graphene-hBN heterostructure, most of the excited surface waves flow to the right side of the source when the chemical potential is low. In contrast, when the chemical potential is high, most of the excited surface waves flow to the left side of the source, as shown in Fig. 7(d).[52] This group-velocity-controlled mechanism was further extended to Janus dipoles in a five-layer graphene-hBN waveguide composed of two weakly coupled graphene-hBN slabs in Fig. 7(e). By tuning the graphene's chemical potential, the system transitions between phonon-like and plasmon-like regimes, as shown in Fig. 7(f).[52] In this configuration, Janus dipoles suppress excitation in one slab while selectively coupling phonon-like or plasmon-like polaritons into the other. Looking forward, leveraging this dynamic group velocity modulation presents a highly promising avenue for actively tailoring the near-field directionality of Janus dipoles, further expanding the toolkit for reconfigurable nanorouting.

## 4. Frontiers and Emerging Trends in Janus Dipoles

In the following we overview emerging research trends which direct ongoing research on Janus dipoles. From the temporal perspective, effort is devoted to introducing relativistic motion to Janus dipoles, investigating their response evolution and mechanisms under dynamic conditions. From the spatial perspective, higher-order multipoles are employed to introduce additional degrees of freedom and enrich near-field manipulation capabilities. At the application level, the intrinsic connection between Janus sources and antenna engineering is being investigated, potentially leading to a new building block for wireless communication, with strong near-field directionality and far-field isotropic radiation. Moreover, establishing connection to the related concept of Janus metasurfaces provides a major growth direction by adapting Janus physics into planar functional architectures, which further link to dynamic, full-space wavefront control by leveraging non-reciprocal or spatial symmetry breaking.

**4.1 Moving Janus sources**

Motion is a fundamental aspect of nature, and the universe itself serves as a grand stage for moving light sources. The light from distant galaxies exhibits cosmological redshift, revealing the expansion of the universe. Supernova explosions eject material at relativistic speeds, their emission encoding the dynamics of these cataclysmic events. Pulsars sweep lighthouse-like beams across the cosmos with extraordinary regularity. These phenomena underscore that motion introduces an additional degree of freedom for the manipulation of light emission and propagation,[111–113] enabling effects far beyond the static regime, particularly when sources approach relativistic velocities.

For dipolar sources, motion profoundly influences their radiation characteristics. The most well-known aspect is the influence on emission frequency, encapsulated in the Doppler effect. Ginzburg and Frank predicted that a source moving faster than the phase velocity of light in the medium ($v \geq v_{\mathrm{p}} = \frac{c}{nv}$, where $n$ is the refractive index) exhibits an anomalous Doppler effect inside the Cherenkov cone, characterized by a transition of the emitted frequency from positive to negative.[114] Recently, this theory was further advanced with the discovery of the superlight inverse Doppler effect,[115] which emerges when the source velocity exceeds twice the phase velocity of light ($v \geq 2v_{\mathrm{p}}$). This effect shows that the received frequency in the direction of motion can be lower than the source frequency even in positive-index systems. The theory was subsequently extended to negative-index systems, where the superlight normal Doppler effect can appear inside the backward Cherenkov cone under the same velocity condition.[116]

Recent investigations into moving sources reveal a diverse range of physical manifestations, which can be effectively illustrated through several representative configurations. A concrete realization of these effects is provided by a moving circularly polarized dipole in Fig. 8(a). In this system, the conventional Doppler effect and superlight inverse Doppler effect manifest on opposite sides of the dipole's trajectory, enabling a clear spatial separation of these distinct frequency shifts.[115] Since each Doppler effect corresponds to a different emission frequency, the radiation direction naturally differs between the two sides. This moving circularly polarized dipole thus serves as a demonstration of how motion can control the frequency and directionality of waves emitted by a dipolar source. In addition to frequency and direction, motion also influences the polarization of dipolar radiation. This is exemplified by a moving Huygens dipole in Fig. 8(b), which was recently predicted to completely suppress the excitation of p-polarized surface plasmons across all Doppler frequencies and directions when traveling parallel to an interface at a critical velocity.[117] Separately from its effects on radiation, motion also alters the fundamental electrodynamic behavior of dipolar sources themselves. A moving dipolar source experiences additional velocity-dependent torque,[118,119] which includes the role played by the hidden dipole moment induced by motion. In addition, a moving dipolar source experiences a self-force due to radiation reaction, interpretable as the momentum carried away by radiation.[120]

Inspired by these well-established phenomena where motion profoundly reshapes the near-field interference of circularly polarized dipoles and Huygens dipoles, a compelling question arises: what new possibilities could emerge when motion is applied to the Janus dipole? While static Janus dipoles have been extensively studied and realized experimentally at both microwave and optical frequencies, their behavior under uniform relativistic motion remains entirely unexplored.

As discussed in previous sections, static Janus dipoles exhibit a unique face-selective coupling governed by the reactive power vector. The velocity-induced phase gradient may modulate the locking between the dipole's internal phase and the reactive power vector, potentially allowing the originally non-coupled face to become partially coupled under specific conditions. The Doppler shift further introduces frequency dependence, raising the possibility of frequency-multiplexed directional coupling where each face dominates at different frequencies. Whether the non-coupling face can maintain its darkness under motion, or whether motion can induce a controlled "leakage" that enables new functionalities, remains an open question.

While the study of moving sources spans from astronomical observations to laboratory demonstrations, the specific case of moving dipolar sources with precisely controlled internal moments remains in its infancy. Unlike astronomical sources whose internal structure is fixed by nature, achieving a specific moving dipolar source requires independent control over the amplitude and phase of both electric and magnetic dipole moments simultaneously, which is a considerably more demanding task. The phenomena discussed above for circularly polarized dipoles and Huygens dipoles represent important theoretical predictions that still await experimental realization. Nevertheless, recent advances in metasurface design, optical trapping, and time-varying photonics offer promising pathways. Techniques developed for generating moving optical spots,[121–123] spatiotemporal interface,[124] and microwave phased dipole arrays[125–127] could be adapted to create moving dipolar sources, as shown in Figs. 8(c, d). Combining these approaches with established experimental realizations of static Janus dipoles could pave the way for the first demonstration of moving Janus sources, unlocking a new frontier in nanophotonic coupling where motion becomes a dynamic knob for controlling near-field directionality.

### 4.2 Toroidal (higher-order) Janus dipoles

The conventional Janus dipole is formed by a pair of orthogonal electric and magnetic dipoles. Beyond the dipole contribution, higher-order multipole interference can introduce additional degrees of freedom and enable more versatile amplitude and phase modulation of optical fields,[60,61] thus facilitating advanced control over directional radiation and waveguiding. Recent studies have proposed a set of higher-order directional dipoles, termed pseudo-directional dipoles (PDDs), where a toroidal dipole replaces the electric dipole component of the conventional circular, Huygens, as well as Janus dipoles.[40,41] The toroidal dipole is induced by circulating poloidal currents on a toroidal surface, exhibiting a more complex charge-current distribution than those of electric or magnetic dipoles. Within the framework of Cartesian multipole expansion, the toroidal dipole moment can be expressed as[128–130]

$$\mathbf{t} = \frac{1}{10c}\int (\mathbf{r}\cdot\mathbf{J})\mathbf{r} - 2r^2\mathbf{J}\mathrm{d}r \tag{4}$$

where $c$ is the speed of light in vacuum, and $\mathbf{J}$ is the poloidal current density. Remarkably, when interacting with free space, the toroidal dipole $\mathbf{t}$ and electric dipole $\mathbf{p}$ belong to the same electric multipole family, with $\mathbf{t}$ being a higher-order term with respect to $\mathbf{p}$.[131] Their radiation fields can cancel with each other under the condition $\mathbf{p} + \mathrm{i}k\mathbf{t} = 0$.[128,129] This forms the basis for the anapole state, a non-radiating charge-current configuration characterized by vanishing far-field radiation.[129,132,133] Moreover, Eq. (4) indicates that the toroidal dipole is odd under both time-reversal and parity transformations. Thus, $\mathrm{i}k\mathbf{t}$ shares the same symmetry properties and near-field patterns with $\mathbf{p}$, allowing the toroidal dipole to replace the electric dipole in the near-field directional coupling and form higher-order directional dipoles.[36,45]

Recalling that the conventional Janus dipole is composed of a pair of orthogonal electric and magnetic dipoles $\pm\pi/2$ out of phase, which can be written as[36,44] $\left(p_i, \pm\mathrm{i}\frac{m_j}{c}\right)$, where $i, j \in x, y, z$, and $i \neq j$. By replacing the electric component $p_i$ with $\mathrm{i}kt_i$, we obtain the expression for the pseudo-Janus dipole[40] $\left(\mathrm{i}kt_i, \pm\mathrm{i}\frac{m_j}{c}\right)$. The pseudo-Janus dipole and associated directional coupling have been demonstrated using two distinct approaches: active sources by magnetic dipoles and passive sources by nanostructures.

For the active realization, the toroidal dipole is equivalent to head-to-tail magnetic dipoles forming a ring.[40] Its amplitude and phase can be flexibly controlled owing to the rich degrees of freedom inherent in the active-magnetic-dipole configuration. Subsequently, one can replace any electric dipole component in the conventional Janus dipole with an active toroidal dipole to achieve analogous face-dependent directional coupling. As illustrated in Fig. 9(a), an active pseudo-Janus dipole can be realized by arranging four active magnetic dipoles head-to-tail into a ring on the $xy$ plane (indicated by a dotted black line), inducing a toroidal dipole $t_z$ along the z direction. An additional active magnetic dipole is placed at the center of the ring to give $\pm\mathrm{i}\frac{m_x}{c}$. When this configuration is positioned between two silicon waveguides as shown in Fig. 9(b), the pseudo-Janus dipole with $t_z = 1/ik$ and $m_x = c$ exhibits face-dependent asymmetric coupling and enhanced directionality tunability through tailoring the source geometry. Notably, the pseudo-Janus dipole can enable switching from bidirectional to unidirectional coupling as well as flipping the directionality without changing the magnetic dipole components, including both head-to-tail and centered magnetic dipoles, which cannot be achieved in conventional systems.[40] In practice, such active pseudo-Janus dipoles are possibly realized using a magnetic dipole antenna combined with poloidal current antennas,[134] paving new ways for designing efficient photonic devices and quantum sources for on-chip applications and optical communications.

Alternatively, passive structures capable of simultaneously exciting toroidal and magnetic dipoles provide another promising route for achieving pseudo-Janus dipoles, which further broaden the application scenarios of higher-order directional dipoles. A viable demonstration is a pair of coupled nano gold helices with opposite handedness as depicted in Fig. 9(c), whose anti-parallel magnetic dipole mode can generate a toroidal dipole.[41,135] By tailoring the internal helix geometry and external illumination, the relative amplitude and phase between induced magnetic, electric, and toroidal dipoles can be tuned, enabling the realization of a passive pseudo-Janus dipole. Under the suitable parameter optimization, the toroidal dipole dominates over the electric dipole and satisfies the pseudo-Janus relationship with the magnetic dipole. Fig. 9(d) illustrates the system configuration for the near-field coupling of the passive pseudo-Janus dipole. Two helices are located at the side of a silicon waveguide under the tilted incidence of a linearly polarized plane wave, as depicted in the upper-left inset. Fig. 9(e) shows face-dependent asymmetric coupling between the helices and the waveguide, confirming that the designed nanostructure can predominantly induce a pseudo-Janus dipole. Analogous to the active pseudo-Janus source, the passive pseudo-Janus structure provides a higher tunability for near-field light manipulation, such as the position-dependent directionality, which is attributed to the complex poloidal current configuration in the helices.[41]

The active and passive realizations of the pseudo-Janus dipole exhibit a higher tunability of the near-field directional coupling, with promising implications across various optical domains. These findings underscore the significant roles of high-order multipoles in on-chip light routing, waveguiding, photonic integrated circuits, non-Hermitian optics, and related applications. The underlying mechanism may be extended to other classical wave systems, such as acoustic waves.

### 4.3 Connections with quasi-isotropic antennas

Isotropic antennas have attracted significant attention in practical applications such as radio frequency identification (RFID),[136–138] wearable communications[139,140] and energy harvesting,[141,142] due to their ability to send wireless signals in all directions. Theoretically, an ideal isotropic antenna radiates electromagnetic waves uniformly throughout the entire spherical space; however, strictly isotropic radiation is physically unrealizable. Consequently, researchers have focused on designing antennas that exhibit nearly isotropic radiation patterns. These antennas are known as quasi-isotropic antennas. It is worth mentioning that gain deviation is the key parameter in evaluating quasi-isotropic performance of an antenna. Gain deviation refers to the difference between the maximum and minimum gain values of the entire radiation sphere. An ideal isotropic antenna will have zero gain deviation. While there is currently no standard definition for the maximum gain deviation of a quasi-isotropic antenna, gain deviations ranging from 2 – 6 dB are commonly achieved in reported quasi-isotropic antennas.[143]

In the development of quasi-isotropic antennas, a method utilizing complementary dipoles has emerged to achieve quasi-isotropic radiation patterns. This approach involves orthogonally placing an electric dipole and a magnetic dipole with a 90° phase difference. The orthogonal placement of the electric and magnetic dipoles ensures that the maximum radiation direction of one dipole aligns with the radiation null of the other as shown in Fig. 10(a).[139] Consequently, the nulls are removed, and the radiation is more evenly spread across all directions. Further, the stipulation of a 90° phase difference mitigates the situation where the radiated fields form the two dipoles interfere significantly, hence creating directions of strong radiation and weak radiation, as in the case of the Huygens source. Theoretically, the Janus source possesses a small gain deviation of only 3 dB.[144] This technique is first applied in[145], where an electric monopole antenna is combined with a pair of rectangular slot antennas to generate the electric and magnetic dipoles, respectively, thereby achieving quasi-isotropic radiation over the upper hemisphere. Subsequently, Ref.[144] employs a dielectric resonator antenna to produce an equivalent magnetic dipole, while a small ground plane generates an electric dipole. The combination yields a quasi-isotropic antenna, as shown in Fig. 10(b). This antenna exhibits a 10-dB impedance bandwidth of 7.3% (2.39 – 2.57 GHz), with a gain variation of 5.6 dB across all radiation directions – exceeding the theoretical value of 3 dB by 2.6 dB. In[146], based on complementary dipoles, a shorted patch antenna is proposed to achieve a quasi-isotropic radiation pattern, as shown in Fig. 10(c). This structure induces a quasi-TEM mode between the upper and lower metal plates, where the magnetic field generates surface electric currents on the shorted sidewalls, and the electric field produces surface magnetic currents at the open aperture. These orthogonal electric and magnetic currents facilitate quasi-isotropic radiation. Experimental results, influenced by practical current distribution discrepancies, shows a gain deviation of 1.95 dB (better than the theoretical 3dB), with a 10-dB impedance bandwidth of 4.48% (2.40 – 2.51 GHz). Driven by the demand for antenna miniaturization, Ref.[147] leverages near-field resonant parasitic (NFRP) elements to design an electrically small antenna (ESA) with $ka < 1$($k$ is the free-space wavenumber, and $a$ is the radius of the minimum circumscribing sphere that encloses the antenna structure.). As shown in Fig. 10(d), the design employs a folded dipole as the electric dipole, and a pair of capacitively loaded loops (CLL) to enhance the magnetic dipole. The antenna produces a nearly isotropic radiation pattern with a gain deviation of 3.14 dB and a 10-dB bandwidth of 0.99% (2.408 – 2.432 GHz).

These antennas consistently employ orthogonal electric and magnetic dipoles with 90° phase shift, which is the same as the physical configuration of a Janus source. However, as antenna engineering primarily prioritizes far-field performance, while most studies on the Janus source emphasize their near-field coupling, the link between these two domains remained largely unexplored until the study in[48], where their connection is experimentally demonstrated. In that work, the quasi-isotropic antenna designed in[147] is positioned between two dielectric waveguides. As shown in Fig. 5(f), the antenna exhibits the characteristic near-field directionality of a Janus source, with the majority of the energy coupling into one side single waveguide. Consequently, the authors call such quasi-isotropic antennas with Janus near-field directionality the Janus antenna. Subsequently, Ref.[55] designs another Janus antenna (Fig. 10(e)), which achieves nearly total suppression of specific polarization components on the uncoupled plane through the coordinated design of the antenna and the upper/lower planar waveguides.

Notwithstanding their physical linkages, the Janus source and quasi-isotropic antenna also share some differences. First, the design methods for quasi-isotropic antennas extend beyond the complementary dipoles to include discrete component arrays[148] and the superposition of multiple dipoles.[149–151] Since these quasi-isotropic antennas do not possess orthogonal electric and magnetic dipole pairs, they do not qualify as Janus antennas. Furthermore, because the directionality of a Janus antenna is confined to the near-field, the physical dimensions of the antenna are strictly constrained to a size that is electrically small. Looking forward, the ongoing advancement of Janus antennas opens up exciting avenues for achieving broader bandwidth, enhanced system integration, and streamlined fabrication techniques, paving the way for their successful transition into practical, real-world applications.

Although the research on Janus antennas is still in its early stage, their potential application scenarios are envisioned to be extensive. In MIMO systems and other array configurations, the cell spacing is typically extremely small. The use of conventional antenna elements in such proximity results in severe mutual coupling, reducing the overall system performance. Janus antennas offer a novel solution[152]: by exhibiting distinct near-field directionality, Janus sources can inherently suppress mutual coupling between elements while maintaining uniform coverage in the far field. The distinctive behavior of Janus antennas – combining near-field directionality with far-field isotropy – enables a high degree of functional integration. They can serve as radiators for external communication, couplers within the system and reduce electromagnetic interference between system components, thereby enabling functional reuse.

### 4.4 Janus metasurfaces

Metasurfaces have provided a powerful paradigm for manipulating electromagnetic (EM) waves by engineering the spatial distribution of subwavelength artificial structures.[153–156] In linear, time-invariant, and passive media, the Lorentz reciprocity theorem imposes a fundamental constraint, typically resulting in symmetric transmission characteristics where the interchange of source and detector yields identical complex transmission coefficients. While such property is inherent to most standard optical components, the demand for increased information capacity and system integration has spurred metasurface research into asymmetric transmission (AT) of EM waves, where wave behavior depends explicitly on the propagation direction (k-vector).[157,158] Generally, the realization of directional metasurface can be broadly categorized into two types with distinct physical mechanisms: non-reciprocal[159] and reciprocal asymmetry.[160] Non-reciprocity fundamentally relies on the breaking of time-reversal symmetry. This is typically achieved by using additional magnetic or non-linear materials or by employing external bias,[161–163] which are often complicated, bulky, and difficult to scale. For example, time-modulated cascaded metasurfaces can achieve non-reciprocal transmission,[164,165] but the transmission efficiency is still a challenging that needs to be enhanced. In contrast, the other type with reciprocal asymmetry operates within the constraint of time-reversal symmetry but relies on the spatial asymmetries, such as mirror or inversion symmetry broken in chiral metastructures.[166] These metasurfaces achieve asymmetric transmission effects because part of the wave is converted to its orthogonal polarizations, while such conversion efficiency is different for the same incidence wave coming from opposite propagation direction.

The EM response of the metasurface can be characterized by the Jones matrix formalism. For a linearly polarized wave propagating in the forward (+z) direction, the transmission matrix $\mathbf{T}^f$ relates the transmitted electric field to the incident fields:

$$\mathbf{T}^f = \begin{pmatrix} t_{xx} & t_{xy} \\ t_{yx} & t_{yy} \end{pmatrix} \tag{5}$$

where $t_{ij}$ denotes the complex transmission coefficient from j-polarization to i-polarization. In the absence of magneto-optical media or active elements, the Lorentz reciprocity theorem governs the system, allowing the backward transmission matrix (−z) to be directly deduced as:

$$\mathbf{T}^b = \begin{pmatrix} t_{xx} & -t_{yx} \\ -t_{xy} & t_{yy} \end{pmatrix} \tag{6}$$

Therefore, in linear and passive reciprocal systems, the co-polarized transmission coefficients remain identical regardless of the propagation direction. Consequently, any directional asymmetry must originate entirely from the off-diagonal cross-polarization components ($t_{xy}$ and $t_{yx}$). By breaking specific spatial symmetries – particularly mirror symmetry along the propagation axis – the structure can be engineered to yield distinct cross-polarization conversion efficiencies for forward and backward waves. The asymmetric transmission parameter is defined as the difference between the total transmitted energies in the forward and backward propagation directions. Specifically, $\Delta^x$ and $\Delta^y$ represent the asymmetric transmission parameters for x- and y-polarized incidence, respectively. Considering the reciprocal relation between the forward and backward transmission matrices, they can be expressed as:

$$\Delta^x = \left|t_{yx}\right|^2 - \left|t_{xy}\right|^2 = -\Delta^y \tag{7}$$

By transforming the basis vectors from linear to circular polarization states, the transmission matrix for CP waves can be directly derived. The matrix, which relates the incident and transmitted waves, is expressed as:

$$\mathbf{T}_{circ}{}^{f} = \begin{pmatrix} t_{++} & t_{+-} \\ t_{-+} & t_{--} \end{pmatrix} = \frac{1}{2}\begin{pmatrix} t_{xx} + t_{yy} + i\left(t_{xy} - t_{yx}\right) & t_{xx} - t_{yy} - i\left(t_{xy} + t_{yx}\right) \\ t_{xx} - t_{yy} + i\left(t_{xy} + t_{yx}\right) & t_{xx} + t_{yy} - i\left(t_{xy} - t_{yx}\right) \end{pmatrix} \quad (8)$$

$$\Delta^{+}{}_{circ} = |t_{-+}|^2 - |t_{+-}|^2 = -\Delta^{-}{}_{circ} \quad (9)$$

where the ‘+’ (‘−’) sign denotes right (left) circular polarization propagating in the +z direction.

Asymmetric transmission can be realized through chiral meta-structures, for instance, fish-scale patterned[157] and multi-layer twisted structures.[167] Moreover, the manipulation of EM waves can be extended from simple asymmetric transmission of energy to more sophisticated wavefront engineering by making the phase response of a metasurface dependent on the incident direction, thus giving rise to directional metasurfaces.[168,169] A key realization of this concept is the Janus metasurface.[170,171] By leveraging the propagation direction as an additional degree of freedom, these metasurfaces facilitate directional wavefront engineering, allowing a single aperture to perform localized EM tasks determined solely by the side of illumination.

The realization of Janus functionalities relies on breaking the symmetric EM responses for opposite propagation directions.[172] As illustrated in Fig. 11(a),[170] this can be achieved by a Janus metasurface element composed of cascaded subwavelength anisotropic impedance sheets. By introducing a rotational twist in geometry to break the out-of-plane symmetry, highly asymmetric transmission is achieved for linearly polarized waves. Furthermore, utilizing field multiplexing to arrange meta-atoms in an interleaved manner allows for independent wavefront manipulation (e.g., free-space holographic imaging shown in the right panel of Fig. 11(a)) for forward and backward propagation directions. Similarly, strategies involving sequential manipulation, such as cascading a gradient metasurface with a polarization-converting element, induce different optical path transformation sequences depending on the incidence direction.[173] For circular polarization control, the Pancharatnam-Berry (PB) phase offers geometric phase modulation,[174–178] while mechanisms that use geometric phases in tandem with propagation or Aharonov-Anandan (AA) phases have been developed to effectively decouple the phase shifts for opposite spin state.[179–181] Besides, optimization approaches utilizing modified Gerchberg-Saxton algorithms have also been employed to numerically optimize phase distributions, enabling – as demonstrated in Fig. 11(b) – the encoding of distinct holographic information for forward and backward waves, effectively circumventing their inherent phase correlations.[182]

While the aforementioned strategies focus primarily on decoupling transmission wavefront phases, extending Janus functionalities to the full EM space requires simultaneous control over reflection and transmission modes. Bidirectional transmission and reflection can be achieved by using meta-atoms with dual capabilities, where the forward and backward metasurfaces are spatially arranged via either partitioning or interleaving. By introducing parametric variations and internal rotations within multilayered meta-atoms, ‘Transmission-Reflection-Integrated’ multiplexing is realized.[171] Fig. 11(c) illustrates a multiplexed Janus metasurface that independently manipulates wavefronts under linearly polarized illuminations with different propagation directions.[183] Similarly, such full-space manipulation can be extended to manipulate circularly polarized waves, where the direction-duplex Janus metasurfaces[184] (Fig. 11(d)) can achieve independent phase control over four circularly polarized channels, enabling full-space quad-channel holography. In addition to spatially multiplexed approaches, receiver-transmitter meta-atom designs – as depicted in Fig. 11(e) – offer a distinct mechanism where the forward and backward transmission phases can be decoupled at the single meta-atom level, enabling the entire aperture to contribute concurrently to bidirectional operations.[185] This configuration supports full-duplex functionality, such as generating distinct multi-focal patterns for opposite propagation directions with high efficiency, thereby maximizing the aperture efficiency. Additionally, as illustrated in Fig. 11(f), to augment information processing capabilities, diffractive neural networks have been integrated into Janus architectures, facilitating simultaneous image transmission and encrypted transformation depending on the propagation direction.[186]

Efforts have also explored integrating active components to realize reconfigurable responses and dynamic wavefront controls.[187] As illustrated in Fig. 11(g), thermally responsive vanadium dioxide ($VO_2$) has been integrated into metasurfaces to realize active Janus architectures in the terahertz regime.[188] $VO_2$ undergoes a reversible insulator-metal-transition (IMT) that alters conductivity and EM response. By leveraging this phase transition, the metasurface can dynamically switch its spatial phase profiles for different incident directions. This capability enables reconfigurable functionalities, such as asymmetric focusing and directional holography, allowing for the manipulation of free-space image projections via temperature control and incident wave direction. Another example in the microwave region is shown in Fig. 11(h), featuring a transparent, conformal device capable of maintaining distinct wavefront manipulations on curved geometries.[189]

The evolution from static asymmetric transmission to dynamic, full-space wavefront manipulation underscores the versatility of Janus metasurfaces in EM wave control. By exploiting propagation direction as an independent dimension, Janus metasurfaces offer an approach to achieve multi-functional integration in optical encryption, holographic displays, and integrated photonic systems.

## Conclusion

The Janus dipole, characterized by its unique combination of strongly directional near-field behavior and quasi-isotropic far-field radiation, has evolved rapidly from a purely theoretical concept into diverse physical realizations spanning optics, microwaves, and acoustics. This review has examined the Janus dipole from three complementary theoretical perspectives – parity symmetry, asymmetric coupling to waveguides, and the electromagnetic wave perspective based on the Poynting vector – providing a comprehensive understanding of its near-field directionality. Significant progress has been made in realizing Janus dipoles in practice. Three main directions have been developed: passive Janus dipoles based on externally excited scatterers, active Janus dipoles that directly generate the required orthogonal dipole moments, and near-field coupling control that enables dynamic modulation of coupling directionality. These realizations have demonstrated not only selective waveguide coupling and mutual coupling suppression but also substantial improvements in efficiency and bandwidth, particularly through the development of 3D active Janus antennas. Looking forward, exciting opportunities lie in the development of tunable and reconfigurable Janus sources, higher-order (e.g., toroidal) multipolar structures and integration with intelligent metasurfaces. Overcoming remaining challenges – such as further efficiency improvement, reconfigurable broadband operation and practical application demonstration – will be essential for realizing the full technological potential of this concept. In summary, the Janus dipole bridges fundamental physics with practical engineering needs by reconciling strong near-field directionality with desirable far-field properties. As research matures, it is poised to become a key enabling technology for next-generation integrated, compact, and multifunctional systems in nanophotonics, antenna engineering, and related fields.

**ACKNOWLEDGMENTS**

This work was supported by the Research Grants Council of Hong Kong, under Grants CityU 11204522 and AoE/E-101/23-N, National Natural Science Foundation of China (No. 12322416), and Research Grants Council of Hong Kong (No. AoE/P-502/20).

**AUTHOR DECLARATIONS Conflict of Interest**

The authors have no conflicts to disclose.

**DATA AVAILABILITY**

Data sharing is not applicable to this article as no new data were created or analyzed in this study.

**REFERENCES**

[1] F.D.M. Haldane, and S. Raghu, "Possible Realization of Directional Optical Waveguides in Photonic Crystals with Broken Time-Reversal Symmetry," Phys. Rev. Lett. **100**(1), 013904 (2008).

[2] J. Petersen, J. Volz, and A. Rauschenbeutel, "Chiral nanophotonic waveguide interface based on spin-orbit interaction of light," Science **346**(6205), 67–71 (2014).

[3] S.V. Li, D.G. Baranov, A.E. Krasnok, and P.A. Belov, "All-dielectric nanoantennas for unidirectional excitation of electromagnetic guided modes," Appl. Phys. Lett. **107**(17), 171101 (2015).

[4] W. Bogaerts, D. Pérez, J. Capmany, D.A.B. Miller, J. Poon, D. Englund, F. Morichetti, and A. Melloni, "Programmable photonic circuits," Nature **586**(7828), 207–216 (2020).

[5] Y. Fang, M. Han, P. Ge, Z. Guo, X. Yu, Y. Deng, C. Wu, Q. Gong, and Y. Liu, "Photoelectronic mapping of the spin–orbit interaction of intense light fields," Nat. Photonics **15**(2), 115–120 (2021).

[6] Z. Yang, Y. Cheng, N. Wang, Y. Chen, and S. Wang, "Nonreciprocal light propagation induced by a subwavelength spinning cylinder," Opt. Express, OE **30**(15), 27993–28002 (2022).

[7] K.Y. Bliokh, D. Smirnova, and F. Nori, "Quantum spin Hall effect of light," Science **348**(6242), 1448–1451 (2015).

[8] C. Sayrin, C. Junge, R. Mitsch, B. Albrecht, D. O'Shea, P. Schneeweiss, J. Volz, and A. Rauschenbeutel, "Nanophotonic Optical Isolator Controlled by the Internal State of Cold Atoms," Phys. Rev. X **5**(4), 041036 (2015).

[9] P. Lodahl, S. Mahmoodian, S. Stobbe, A. Rauschenbeutel, P. Schneeweiss, J. Volz, H. Pichler,

and P. Zoller, “Chiral quantum optics,” Nature **541**(7638), 473–480 (2017).
[10] A. Kurs, A. Karalis, R. Moffatt, J.D. Joannopoulos, P. Fisher, and M. Soljačić, “Wireless Power Transfer via Strongly Coupled Magnetic Resonances,” Science **317**(5834), 83–86 (2007).
[11] S. Assawaworrarit, X. Yu, and S. Fan, “Robust wireless power transfer using a nonlinear parity–time-symmetric circuit,” Nature **546**(7658), 387–390 (2017).
[12] L. Zhu, A. Fiorino, D. Thompson, R. Mittapally, E. Meyhofer, and P. Reddy, “Near-field photonic cooling through control of the chemical potential of photons,” Nature **566**(7743), 239–244 (2019).
[13] M. Neugebauer, P. Woźniak, A. Bag, G. Leuchs, and P. Banzer, “Polarization-controlled directional scattering for nanoscopic position sensing,” Nat Commun **7**(1), 11286 (2016).
[14] A. Bag, M. Neugebauer, U. Mick, S. Christiansen, S.A. Schulz, and P. Banzer, “Towards fully integrated photonic displacement sensors,” Nature Communications **11**(1), 1–7 (2020).
[15] T. Zang, H. Zang, Z. Xi, J. Du, H. Wang, Y. Lu, and P. Wang, “Asymmetric Excitation of Surface Plasmon Polaritons via Paired Slot Antennas for Angstrom Displacement Sensing,” Phys. Rev. Lett. **124**(24), 243901 (2020).
[16] R. Zhu, C. Qian, S. Xiao, J. Yang, S. Yan, H. Liu, D. Dai, H. Li, L. Yang, X. Chen, Y. Yuan, D. Dai, Z. Zuo, H. Ni, Z. Niu, C. Wang, K. Jin, Q. Gong, and X. Xu, “Full polarization control of photons with evanescent wave coupling in the ultra subwavelength gap of photonic molecules,” Light Sci Appl **14**(1), 114 (2025).
[17] F.J. Rodríguez-Fortuño, G. Marino, P. Ginzburg, D. O’Connor, A. Martínez, G.A. Wurtz, and A.V. Zayats, “Near-Field Interference for the Unidirectional Excitation of Electromagnetic Guided Modes,” Science **340**(6130), 328–330 (2013).
[18] D. O’Connor, P. Ginzburg, F.J. Rodríguez-Fortuño, G.A. Wurtz, and A.V. Zayats, “Spin–orbit coupling in surface plasmon scattering by nanostructures,” Nat Commun **5**(1), 5327 (2014).
[19] B. le Feber, N. Rotenberg, and L. Kuipers, “Nanophotonic control of circular dipole emission,” Nat Commun **6**(1), 6695 (2015).
[20] K.Y. Bliokh, F.J. Rodríguez-Fortuño, F. Nori, and A.V. Zayats, “Spin–orbit interactions of light,” Nature Photon **9**(12), 796–808 (2015).
[21] A. Espinosa-Soria, and A. Martínez, “Transverse Spin and Spin-Orbit Coupling in Silicon Waveguides,” IEEE Photonics Technology Letters **28**(14), 1561–1564 (2016).
[22] M.F. Picardi, A. Manjavacas, A.V. Zayats, and F.J. Rodríguez-Fortuño, “Unidirectional evanescent-wave coupling from circularly polarized electric and magnetic dipoles: An angular spectrum approach,” Phys. Rev. B **95**(24), 245416 (2017).
[23] P. Shi, X. Lei, Q. Zhang, H. Li, L. Du, and X. Yuan, “Intrinsic Spin-Momentum Dynamics of Surface Electromagnetic Waves in Dispersive Interfaces,” Phys. Rev. Lett. **128**(21), 213904 (2022).
[24] S. Wang, B. Hou, W. Lu, Y. Chen, Z.Q. Zhang, and C.T. Chan, “Arbitrary order exceptional point induced by photonic spin–orbit interaction in coupled resonators,” Nat Commun **10**(1), 832 (2019).
[25] P. Jin, and R.W. Ziolkowski, “Metamaterial-Inspired, Electrically Small Huygens Sources,” IEEE Antennas and Wireless Propagation Letters **9**, 501–505 (2010).
[26] A.B. Evlyukhin, S.M. Novikov, U. Zywietz, R.L. Eriksen, C. Reinhardt, S.I. Bozhevolnyi, and B.N. Chichkov, “Demonstration of Magnetic Dipole Resonances of Dielectric Nanospheres in the Visible Region,” Nano Lett. **12**(7), 3749–3755 (2012).
[27] J.M. Geffrin, B. García-Cámara, R. Gómez-Medina, P. Albella, L.S. Froufe-Pérez, C. Eyraud, A. Litman, R. Vaillon, F. González, M. Nieto-Vesperinas, J.J. Sáenz, and F. Moreno, “Magnetic and electric coherence in forward- and back-scattered electromagnetic waves by a single dielectric

subwavelength sphere," Nat Commun **3**(1), 1171 (2012).
[28] Y.H. Fu, A.I. Kuznetsov, A.E. Miroshnichenko, Y.F. Yu, and B. Luk'yanchuk, "Directional visible light scattering by silicon nanoparticles," Nat Commun **4**(1), 1527 (2013).
[29] C. Pfeiffer, and A. Grbic, "Metamaterial Huygens' Surfaces: Tailoring Wave Fronts with Reflectionless Sheets," Phys. Rev. Lett. **110**(19), 197401 (2013).
[30] T. Coenen, F. Bernal Arango, A. Femius Koenderink, and A. Polman, "Directional emission from a single plasmonic scatterer," Nat Commun **5**(1), 3250 (2014).
[31] D. Permyakov, I. Sinev, D. Markovich, P. Ginzburg, A. Samusev, P. Belov, V. Valuckas, A.I. Kuznetsov, B.S. Luk'yanchuk, A.E. Miroshnichenko, D.N. Neshev, and Y.S. Kivshar, "Probing magnetic and electric optical responses of silicon nanoparticles," Appl. Phys. Lett. **106**(17), 171110 (2015).
[32] K. Yao, and Y. Liu, "Controlling Electric and Magnetic Resonances for Ultracompact Nanoantennas with Tunable Directionality," ACS Photonics **3**(6), 953–963 (2016).
[33] A.I. Kuznetsov, A.E. Miroshnichenko, M.L. Brongersma, Y.S. Kivshar, and B. Luk'yanchuk, "Optically resonant dielectric nanostructures," Science **354**(6314), aag2472 (2016).
[34] S. Nechayev, J.S. Eismann, M. Neugebauer, P. Woźniak, A. Bag, G. Leuchs, and P. Banzer, "Huygens' dipole for polarization-controlled nanoscale light routing," Physical Review A **99**(4), 041801 (2019).
[35] Y. Zhong, C. Wang, C. Bian, X. Chen, J. Chen, X. Zhu, H. Hu, T. Low, H. Chen, B. Zhang, and X. Lin, "Near-field directionality governed by asymmetric dipole–matter interactions," Opt. Lett., OL **49**(4), 826–829 (2024).
[36] M.F. Picardi, A.V. Zayats, and F.J. Rodríguez-Fortuño, "Janus and Huygens Dipoles: Near-Field Directionality Beyond Spin-Momentum Locking," Phys. Rev. Lett. **120**(11), 117402 (2018).
[37] M. Picardi, A. Zayats, and F. Rodríguez-Fortuño, "Not every dipole is the same: The hidden patterns of dipolar near fields," Europhysics News **49**(4), 14–18 (2018).
[38] L. Wei, and F.J. Rodríguez-Fortuño, "Momentum-Space Geometric Structure of Helical Evanescent Waves and Its Implications on Near-Field Directionality," Phys. Rev. Applied **13**(1), 014008 (2020).
[39] H. Jiang, J. Wang, G. Song, J. Ren, X. Yang, J. Xu, and Y. Yang, "Near-field Properties of Spin, Huygens and Janus Sources in a Narrow Sandwiched Structure," J. Phys. B: At. Mol. Opt. Phys. **55**(15), 155001 (2022).
[40] J. Jung, Y. Cheng, W. Xiao, and S. Wang, "Directional sources realized by toroidal dipoles," Phys. Rev. A **110**(6), 063510 (2024).
[41] J. Jung, Y. Cheng, and S. Wang, "Toroidal-dipole-assisted realization of directional sources by coupled chiral particles," Opt. Express, OE **33**(23), 49551–49561 (2025).
[42] T. Wu, A. Baron, P. Lalanne, and K. Vynck, "Intrinsic multipolar contents of nanoresonators for tailored scattering," Phys. Rev. A **101**(1), 011803 (2020).
[43] L. Wei, and F.J. Rodríguez-Fortuño, "Far-field and near-field directionality in acoustic scattering," New J. Phys. **22**(8), 083016 (2020).
[44] Y. Cheng, K.A. Oyesina, B. Xue, D. Lei, A.M.H. Wong, and S. Wang, "Directional dipole dice enabled by anisotropic chirality," Proceedings of the National Academy of Sciences **120**(25), e2301620120 (2023).
[45] Y. Long, J. Ren, Z. Guo, H. Jiang, Y. Wang, Y. Sun, and H. Chen, "Designing all-electric subwavelength metasources for near-field photonic routings," Physical Review Letters **125**(15), 157401 (2020).
[46] Y. Long, H. Ge, D. Zhang, X. Xu, J. Ren, M.-H. Lu, M. Bao, H. Chen, and Y.-F. Chen,

"Symmetry selective directionality in near-field acoustics," Natl Sci Rev **7**(6), 1024–1035 (2020).
[47] B. Xue, K.A. Oyesina, and A.M.H. Wong, "Electromagnetic near-field mutual coupling suppression with active Janus sources," Commun Phys **7**(1), 1–9 (2024).
[48] B. Xue, K.A. Oyesina, and A.M.H. Wong, "Achieving Broadband Near-Field Directionality with a 3D Active Janus Antenna," Laser & Photonics Reviews **19**(2), 2401093 (2025).
[49] M.F. Picardi, M. Neugebauer, J.S. Eismann, G. Leuchs, P. Banzer, F.J. Rodríguez-Fortuño, and A.V. Zayats, "Experimental demonstration of linear and spinning Janus dipoles for polarisation- and wavelength-selective near-field coupling," Light Sci Appl **8**(1), 52 (2019).
[50] M.F. Picardi, C.P.T. McPolin, J.J. Kingsley-Smith, X. Zhang, S. Xiao, F.J. Rodríguez-Fortuño, and A.V. Zayats, "Integrated Janus dipole source for selective coupling to silicon waveguide networks," Applied Physics Reviews **9**(2), 021410 (2022).
[51] C. Wang, Y. Zhong, X. Chen, H. Wang, T. Low, H. Chen, B. Zhang, and X. Lin, "Near-Field Coupling of Janus Dipoles Beyond Polarization Locking," Laser & Photonics Reviews **n/a**(n/a), 2301035 (2024).
[52] Y. Jiang, X. Lin, T. Low, B. Zhang, and H. Chen, "Group-Velocity-Controlled and Gate-Tunable Directional Excitation of Polaritons in Graphene-Boron Nitride Heterostructures," Laser & Photonics Reviews **12**(5), 1800049 (2018).
[53] Y. Jiang, X. Lin, and and H. Chen, "Directional Polaritonic Excitation of Circular, Huygens and Janus Dipoles in Graphene-Hexagonal Boron Nitride Heterostructures," Progress In Electromagnetics Research **170**, 169–176 (2021).
[54] Y. Zhong, X. Lin, J. Jiang, Y. Yang, G.-G. Liu, H. Xue, T. Low, H. Chen, and B. Zhang, "Toggling Near-Field Directionality via Polarization Control of Surface Waves," Laser & Photonics Reviews **15**(4), 2000388 (2021).
[55] W. Ma, X. Chen, Y. Zhong, C. Bian, G. Zhao, C. Wang, H. Chen, and X. Lin, "Dual-polarization perfect Janus dipole in near-field coupling," Phys. Rev. A **112**(4), (2025).
[56] Y. Cheng, Y.-S. Zeng, W. Xiao, T. Fu, J. Wu, G.-B. Wu, D.P. Tsai, and S. Wang, "Riemann-Silberstein geometric phase in 4D polarization space," arXiv.Org, (2025).
[57] A. Aiello, P. Banzer, M. Neugebauer, and G. Leuchs, "From transverse angular momentum to photonic wheels," Nature Photon **9**(12), 789–795 (2015).
[58] T.V. Mechelen, and Z. Jacob, "Universal spin-momentum locking of evanescent waves," Optica, OPTICA **3**(2), 118–126 (2016).
[59] P. Shi, L. Du, C. Li, A.V. Zayats, and X. Yuan, "Transverse spin dynamics in structured electromagnetic guided waves," Proceedings of the National Academy of Sciences **118**(6), (2021).
[60] M. F. Picardi, A. V. Zayats, and F. J. Rodríguez-Fortuño, "Amplitude and Phase Control of Guided Modes Excitation from a Single Dipole Source: Engineering Far-and Near-Field Directionality," Laser & Photonics Reviews **13**(12), 1900250 (2019).
[61] J.E. Vázquez-Lozano, A. Martínez, and F.J. Rodríguez-Fortuño, "Near-field directionality beyond the dipole approximation: electric quadrupole and higher-order multipole angular spectra," Physical Review Applied **12**(2), 024065 (2019).
[62] J. Jiang, and Q.-X. Chu, "Broadband Decoupling for Antenna Arrays Using Multiple Decoupling Nulls," IEEE Transactions on Antennas and Propagation **71**(11), 8616–8627 (2023).
[63] M. Li, B.G. Zhong, and S.W. Cheung, "Isolation Enhancement for MIMO Patch Antennas Using Near-Field Resonators as Coupling-Mode Transducers," IEEE Transactions on Antennas and Propagation **67**(2), 755–764 (2019).
[64] J. Weber, C. Volmer, K. Blau, R. Stephan, and M.A. Hein, "Miniaturized antenna arrays using decoupling networks with realistic elements," IEEE Transactions on Microwave Theory and

Techniques **54**(6), 2733–2740 (2006).
[65] J. Sui, and K.-L. Wu, “Self-Curing Decoupling Technique for Two Inverted-F Antennas With Capacitive Loads,” IEEE Transactions on Antennas and Propagation **66**(3), 1093–1101 (2018).
[66] K.-L. Wong, M.-F. Jian, C.-J. Chen, and J.-Z. Chen, “Two-Port Same-Polarized Patch Antenna Based on Two Out-of-Phase TM10 Modes for Access-Point MIMO Antenna Application,” IEEE Antennas and Wireless Propagation Letters **20**(4), 572–576 (2021).
[67] J. Xu, X. He, and T. Deng, “A Self-Decoupled MIMO Patch Array With Consistent Radiation Patterns,” IEEE Transactions on Antennas and Propagation **72**(12), 8971–8979 (2024).
[68] B. Kannan, A. Almanakly, Y. Sung, A. Di Paolo, D.A. Rower, J. Braumüller, A. Melville, B.M. Niedzielski, A. Karamlou, K. Serniak, A. Vepsäläinen, M.E. Schwartz, J.L. Yoder, R. Winik, J.I.-J. Wang, T.P. Orlando, S. Gustavsson, J.A. Grover, and W.D. Oliver, “On-demand directional microwave photon emission using waveguide quantum electrodynamics,” Nat. Phys. **19**(3), 394–400 (2023).
[69] T.J. Cui, S. Zhang, A. Alù, M. Wegener, S.J. Pendry, J. Luo, Y. Lai, Z. Wang, X. Lin, H. Chen, P. Chen, R.-X. Wu, Y. Yin, P. Zhao, H. Chen, Y. Li, Z. Zhou, N. Engheta, V. Asadchy, C. Simovski, S. Tretyakov, B. Yang, S.D. Campbell, Y. Hao, D.H. Werner, S. Sun, L. Zhou, S. Xu, H.-B. Sun, Z. Zhou, Z. Li, G. Zheng, X. Chen, T. Li, S. Zhu, J. Zhou, J. Zhao, Z. Liu, Y. Zhang, Q. Zhang, M. Gu, S. Xiao, Y. Liu, X. Zhang, Y. Tang, G. Li, T. Zentgraf, K. Koshelev, Y. Kivshar, X. Li, T. Badloe, L. Huang, J. Rho, S. Wang, D.P. Tsai, A.Y. Bykov, A.V. Krasavin, A.V. Zayats, C. McDonnell, T. Ellenbogen, X. Luo, M. Pu, F.J. Garcia-Vidal, L. Liu, Z. Li, W. Tang, H.F. Ma, J. Zhang, Y. Luo, X. Zhang, H.C. Zhang, P.H. He, L.P. Zhang, X. Wan, H. Wu, S. Liu, W.X. Jiang, X.G. Zhang, C.-W. Qiu, Q. Ma, C. Liu, L. Li, J. Han, L. Li, M. Cotrufo, C. Caloz, Z.-L. Deck-Léger, A. Bahrami, O. Céspedes, E. Galiffi, P.A. Huidobro, Q. Cheng, J.Y. Dai, J.C. Ke, L. Zhang, V. Galdi, and M. di Renzo, “Roadmap on electromagnetic metamaterials and metasurfaces,” J. Phys. Photonics **6**(3), 032502 (2024).
[70] M.V. Larsen, J.E. Bourassa, S. Kocsis, J.F. Tasker, R.S. Chadwick, C. González-Arciniegas, J. Hastrup, C.E. Lopetegui-González, F.M. Miatto, A. Motamedi, R. Noro, G. Roeland, R. Baby, H. Chen, P. Contu, I. Di Luch, C. Drago, M. Giesbrecht, T. Grainge, I. Krasnokutska, M. Menotti, B. Morrison, C. Puviraj, K. Rezaei Shad, B. Hussain, J. McMahon, J.E. Ortmann, M.J. Collins, C. Ma, D.S. Phillips, M. Seymour, Q.Y. Tang, B. Yang, Z. Vernon, R.N. Alexander, and D.H. Mahler, “Integrated photonic source of Gottesman–Kitaev–Preskill qubits,” Nature **642**(8068), 587–591 (2025).
[71] D. Choi, K.Y. Lee, D.-J. Shin, J.W. Yoon, and S.-H. Gong, “Unidirectional guided resonance continuum of Dirac bands in WS2 bilayer metasurfaces,” Nat. Nanotechnol. **20**(8), 1026–1033 (2025).
[72] Y. Zhong, T. Cai, T. Low, H. Chen, and X. Lin, “Optical interface engineering with on-demand magnetic surface conductivities,” Phys. Rev. B **106**(3), 035304 (2022).
[73] C. Bian, X. Zhang, W. Ma, X. Chen, H. Chen, T. Low, and X. Lin, “Antihyperbolic surface waves on hyperbolic metasurfaces,” Phys. Rev. A **111**(3), 033522 (2025).
[74] C. Bian, Y. Zhong, X. Chen, T. Low, H. Chen, B. Zhang, and X. Lin, “Janus faces of dipolar sources in directional near-field coupling with an oriented misalignment,” Phys. Rev. A **109**(3), 033505 (2024).
[75] Q. Zhang, G. Hu, W. Ma, P. Li, A. Krasnok, R. Hillenbrand, A. Alù, and C.-W. Qiu, “Interface nano-optics with van der Waals polaritons,” Nature **597**(7875), 187–195 (2021).
[76] A. Kavokin, T.C.H. Liew, C. Schneider, P.G. Lagoudakis, S. Klembt, and S. Hoefling, “Polariton condensates for classical and quantum computing,” Nat Rev Phys **4**(7), 435–451 (2022).

[77] T. Cai, Y. Zhong, D. Liu, H. Huang, D. Wang, Y. Yang, H. Chen, and X. Lin, "Observation of Polarization-maintaining Near-field Directionality," PIER **181**, 35–41 (2024).
[78] J. Peng, R.-Y. Zhang, S. Jia, W. Liu, and S. Wang, "Topological near fields generated by topological structures," Science Advances **8**(41), eabq0910 (2022).
[79] T. Fu, R.-Y. Zhang, S. Jia, C.T. Chan, and S. Wang, "Near-Field Spin Chern Number Quantized by Real-Space Topology of Optical Structures," Phys. Rev. Lett. **132**(23), 233801 (2024).
[80] T. Fu, Q. Tong, S. Jia, and S. Wang, "Topological Dark Spots of the Electric Near Field in Metal Structures," ACS Photonics **11**(10), 4342–4348 (2024).
[81] T.L. Cocker, V. Jelic, R. Hillenbrand, and F.A. Hegmann, "Nanoscale terahertz scanning probe microscopy," Nat. Photon. **15**(8), 558–569 (2021).
[82] Y. Tang, and A.E. Cohen, "Optical Chirality and Its Interaction with Matter," Phys. Rev. Lett. **104**(16), 163901 (2010).
[83] J.S. Eismann, M. Neugebauer, and P. Banzer, "Exciting a chiral dipole moment in an achiral nanostructure," Optica **5**(8), 954–959 (2018).
[84] S.-H. Gong, F. Alpeggiani, B. Sciacca, E.C. Garnett, and L. Kuipers, "Nanoscale chiral valley-photon interface through optical spin-orbit coupling," Science **359**(6374), 443–447 (2018).
[85] J.S. Lee, N. Farmakidis, C.D. Wright, and H. Bhaskaran, "Polarization-selective reconfigurability in hybridized-active-dielectric nanowires," Science Advances **8**(24), eabn9459 (2022).
[86] P.V. Kapitanova, P. Ginzburg, F.J. Rodríguez-Fortuño, D.S. Filonov, P.M. Voroshilov, P.A. Belov, A.N. Poddubny, Y.S. Kivshar, G.A. Wurtz, and A.V. Zayats, "Photonic spin Hall effect in hyperbolic metamaterials for polarization-controlled routing of subwavelength modes," Nat Commun **5**(1), 3226 (2014).
[87] Z. Xu, J. Chang, J. Tong, D.F. Sievenpiper, and T.J. Cui, "Near-field chiral excitation of universal spin-momentum locking transport of edge waves in microwave metamaterials," AP **4**(4), 046004 (2022).
[88] S.D. Namgung, R.M. Kim, Y.-C. Lim, J.W. Lee, N.H. Cho, H. Kim, J.-S. Huh, H. Rhee, S. Nah, M.-K. Song, J.-Y. Kwon, and K.T. Nam, "Circularly polarized light-sensitive, hot electron transistor with chiral plasmonic nanoparticles," Nat Commun **13**(1), 5081 (2022).
[89] J. Ni, S. Ji, Z. Wang, S. Liu, Y. Hu, Y. Chen, J. Li, X. Li, J. Chu, D. Wu, and C.-W. Qiu, "Unidirectional unpolarized luminescence emission via vortex excitation," Nat. Photon. **17**(7), 601–606 (2023).
[90] D.V. Zhirihin, M.S. Sidorenko, A.D. Rozenblit, G.D. Kurganov, M.A. Gorlach, D.S. Filonov, Y.S. Kivshar, and A.P. Slobozhanyuk, "Helical metasurfaces based on topological surface states in three-dimensional photonic topological insulators," Nat. Mater. **25**(5), 762–766 (2026).
[91] Y. Cheng, W. Xiao, and S. Wang, "Multiplexing spectral line shape of waveguide transmission by photonic spin-orbit interaction," Phys. Rev. A **108**(4), 043511 (2023).
[92] S. Kim, Y.-C. Lim, R.M. Kim, J.E. Fröch, T.N. Tran, K.T. Nam, and I. Aharonovich, "A Single Chiral Nanoparticle Induced Valley Polarization Enhancement," Small **16**(37), 2003005 (2020).
[93] Y. Xie, A.V. Krasavin, D.J. Roth, and A.V. Zayats, "Unidirectional chiral scattering from single enantiomeric plasmonic nanoparticles," Nat Commun **16**(1), 1125 (2025).
[94] Y. Chen, Y. Chen, Y. Fang, R. Ai, X. Cui, X. Zhuo, and J. Wang, "Photonic spin-Hall effect in chiral plasmonic assemblies," Nat Commun **17**(1), 3246 (2026).
[95] S.B. Wang, and C.T. Chan, "Lateral optical force on chiral particles near a surface," Nat Commun **5**(1), 3307 (2014).
[96] Y. Liang, K. Koshelev, F. Zhang, H. Lin, S. Lin, J. Wu, B. Jia, and Y. Kivshar, "Bound States in

the Continuum in Anisotropic Plasmonic Metasurfaces," Nano Lett. **20**(9), 6351–6356 (2020).
[97] O. Yermakov, V. Lenets, A. Sayanskiy, J. Baena, E. Martini, S. Glybovski, and S. Maci, "Surface Waves on Self-Complementary Metasurfaces: All-Frequency Hyperbolicity, Extreme Canalization, and TE-TM Polarization Degeneracy," Phys. Rev. X **11**(3), 031038 (2021).
[98] X. Zhang, J. Chen, R. Chen, C. Wang, T. Cai, R. Abdi-Ghaleh, H. Chen, and X. Lin, "Perspective on Meta-Boundaries," ACS Photonics **10**(7), 2102–2115 (2023).
[99] X. Lin, Z. Liu, T. Stauber, G. Gómez-Santos, F. Gao, H. Chen, B. Zhang, and T. Low, "Chiral Plasmons with Twisted Atomic Bilayers," Phys. Rev. Lett. **125**(7), 077401 (2020).
[100] J. Cai, W. Zhang, L. Xu, C. Hao, W. Ma, M. Sun, X. Wu, X. Qin, F.M. Colombari, A.F. de Moura, J. Xu, M.C. Silva, E.B. Carneiro-Neto, W.R. Gomes, R.A.L. Vallée, E.C. Pereira, X. Liu, C. Xu, R. Klajn, N.A. Kotov, and H. Kuang, "Polarization-sensitive optoionic membranes from chiral plasmonic nanoparticles," Nat. Nanotechnol. **17**(4), 408–416 (2022).
[101] T. Huang, X. Tu, C. Shen, B. Zheng, J. Wang, H. Wang, K. Khaliji, S.H. Park, Z. Liu, T. Yang, Z. Zhang, L. Shao, X. Li, T. Low, Y. Shi, and X. Wang, "Observation of chiral and slow plasmons in twisted bilayer graphene," Nature **605**(7908), 63–68 (2022).
[102] O. Takayama, D. Artigas, and L. Torner, "Lossless directional guiding of light in dielectric nanosheets using Dyakonov surface waves," Nature Nanotech **9**(6), 419–424 (2014).
[103] H. Hu, X. Lin, L.J. Wong, Q. Yang, D. Liu, B. Zhang, and Y. Luo, "Surface Dyakonov–Cherenkov radiation," eLight **2**(1), 2 (2022).
[104] G. Hu, W. Ma, D. Hu, J. Wu, C. Zheng, K. Liu, X. Zhang, X. Ni, J. Chen, X. Zhang, Q. Dai, J.D. Caldwell, A. Paarmann, A. Alù, P. Li, and C.-W. Qiu, "Real-space nanoimaging of hyperbolic shear polaritons in a monoclinic crystal," Nat. Nanotechnol. **18**(1), 64–70 (2023).
[105] W. Liu, J. Liu, W. Wu, L. Qu, Y. Zhang, X. Liu, C. Zhu, C. Wang, W. Cai, M. Ren, and J. Xu, "Electro-Optic Lithium Niobate Metasurface Enabling Dynamic Polarization Control," ACS Photonics **13**(7), 1767–1773 (2026).
[106] A. Woessner, M.B. Lundeberg, Y. Gao, A. Principi, P. Alonso-González, M. Carrega, K. Watanabe, T. Taniguchi, G. Vignale, M. Polini, J. Hone, R. Hillenbrand, and F.H.L. Koppens, "Highly confined low-loss plasmons in graphene–boron nitride heterostructures," Nature Mater **14**(4), 421–425 (2015).
[107] S. Dai, Q. Ma, T. Andersen, A.S. Mcleod, Z. Fei, M.K. Liu, M. Wagner, K. Watanabe, T. Taniguchi, M. Thiemens, F. Keilmann, P. Jarillo-Herrero, M.M. Fogler, and D.N. Basov, "Subdiffractional focusing and guiding of polaritonic rays in a natural hyperbolic material," Nat Commun **6**(1), 6963 (2015).
[108] A. Kumar, T. Low, K.H. Fung, P. Avouris, and N.X. Fang, "Tunable Light–Matter Interaction and the Role of Hyperbolicity in Graphene–hBN System," Nano Lett. **15**(5), 3172–3180 (2015).
[109] W. Hutchins, S. Zare, D.M. Hirt, J.A. Tomko, J.R. Matson, K. Diaz-Granados, M. Long, M. He, T. Pfeifer, J. Li, J.H. Edgar, J.-P. Maria, J.D. Caldwell, and P.E. Hopkins, "Ultrafast evanescent heat transfer across solid interfaces via hyperbolic phonon–polariton modes in hexagonal boron nitride," Nat. Mater. **24**(5), 698–706 (2025).
[110] M. Trushin, "Electrical Generation of Surface Plasmon Polaritons in Plasmonic Heterostructures," Phys. Rev. Lett. **136**(1), 016901 (2026).
[111] Z. Yang, W. Xiao, H. Li, H. Pan, and S. Wang, "Nonreciprocal optical metasurface based on spinning cylinders," Phys. Rev. A **112**(1), 013519 (2025).
[112] H. Shi, Y. Cheng, Z. Yang, Y. Chen, and S. Wang, "Optical isolation induced by subwavelength spinning particle via spin-orbit interaction," Phys. Rev. B **103**(9), 094105 (2021).
[113] R. Huang, A. Miranowicz, J.-Q. Liao, F. Nori, and H. Jing, "Nonreciprocal Photon Blockade,"

Phys. Rev. Lett. **121**(15), 153601 (2018).
[114] V.L. Ginzburg, and I.M. Frank, "On the Doppler effect at the superluminal velocity," in *Dokl. Akad. Nauk SSSR*, (1947), pp. 583–586.
[115] X. Shi, X. Lin, I. Kaminer, F. Gao, Z. Yang, J.D. Joannopoulos, M. Soljačić, and B. Zhang, "Superlight inverse Doppler effect," Nature Phys **14**(10), 1001–1005 (2018).
[116] X. Lin, and B. Zhang, "Normal Doppler Frequency Shift in Negative Refractive-Index Systems," Laser & Photonics Reviews **13**(12), 1900081 (2019).
[117] X. Chen, Y. Zhong, W. Ma, B. Zhang, H. Chen, T. Low, and X. Lin, "Critical polarization suppression in the near-field interference of moving Huygens-like dipoles," Opt. Lett., OL **50**(14), 4562–4565 (2025).
[118] A.L. Kholmetskii, O.V. Missevitch, and T. Yarman, "Torque on a moving electric/magnetic dipole," Progress In Electromagnetics Research B **45**, 83–99 (2012).
[119] A. Kholmetskii, O. Missevitch, and T. Yarman, "Electric/magnetic dipolein an electromagnetic field: force, torque and energy," Eur. Phys. J. Plus **129**(10), 215 (2014).
[120] K.A. Milton, H. Day, Y. Li, X. Guo, and G. Kennedy, "Self-force on moving electric and magnetic dipoles: Dipole radiation, Vavilov-\ifmmode \check{C}\else \v{C}\fi{}erenkov radiation, friction with a conducting surface, and the Einstein-Hopf effect," Phys. Rev. Res. **2**(4), 043347 (2020).
[121] B.C. Smith, J.F. Whitaker, and S.C. Rand, "Steerable THz pulses from thin emitters via optical pulse-front tilt," Opt. Express, OE **24**(18), 20755–20762 (2016).
[122] M.I. Bakunov, M.V. Tsarev, and M. Hangyo, "Cherenkov emission of terahertz surface plasmon polaritons from a superluminal optical spot on a structured metal surface," Opt. Express, OE **17**(11), 9323–9329 (2009).
[123] D.H. Auston, K.P. Cheung, J.A. Valdmanis, and D.A. Kleinman, "Cherenkov Radiation from Femtosecond Optical Pulses in Electro-Optic Media," Phys. Rev. Lett. **53**(16), 1555–1558 (1984).
[124] D. Oue, K. Ding, and J.B. Pendry, "Čerenkov radiation in vacuum from a superluminal grating," Phys. Rev. Research **4**(1), 013064 (2022).
[125] L. Jing, X. Lin, Z. Wang, I. Kaminer, H. Hu, E. Li, Y. Liu, M. Chen, B. Zhang, and H. Chen, "Polarization Shaping of Free-Electron Radiation by Gradient Bianisotropic Metasurfaces," Laser & Photonics Reviews **15**(4), 2000426 (2021).
[126] P. Genevet, D. Wintz, A. Ambrosio, A. She, R. Blanchard, and F. Capasso, "Controlled steering of Cherenkov surface plasmon wakes with a one-dimensional metamaterial," Nature Nanotech **10**(9), 804–809 (2015).
[127] S. Xi, H. Chen, T. Jiang, L. Ran, J. Huangfu, B.-I. Wu, J.A. Kong, and M. Chen, "Experimental Verification of Reversed Cherenkov Radiation in Left-Handed Metamaterial," Phys. Rev. Lett. **103**(19), 194801 (2009).
[128] X.-L. Zhang, S.B. Wang, Z. Lin, H.-B. Sun, and C.T. Chan, "Optical force on toroidal nanostructures: Toroidal dipole versus renormalized electric dipole," Phys. Rev. A **92**(4), 043804 (2015).
[129] A.E. Miroshnichenko, A.B. Evlyukhin, Y.F. Yu, R.M. Bakker, A. Chipouline, A.I. Kuznetsov, B. Luk'yanchuk, B.N. Chichkov, and Y.S. Kivshar, "Nonradiating anapole modes in dielectric nanoparticles," Nat Commun **6**(1), 8069 (2015).
[130] E.E. Radescu, and G. Vaman, "Exact calculation of the angular momentum loss, recoil force, and radiation intensity for an arbitrary source in terms of electric, magnetic, and toroid multipoles," Phys. Rev. E **65**(4), 046609 (2002).
[131] R. Alaee, C. Rockstuhl, and I. Fernandez-Corbaton, "An electromagnetic multipole expansion

beyond the long-wavelength approximation," Optics Communications **407**, 17–21 (2018).
[132] V.A. Zenin, A.B. Evlyukhin, S.M. Novikov, Y. Yang, R. Malureanu, A.V. Lavrinenko, B.N. Chichkov, and S.I. Bozhevolnyi, "Direct Amplitude-Phase Near-Field Observation of Higher-Order Anapole States," Nano Lett. **17**(11), 7152–7159 (2017).
[133] A.A. Basharin, V. Chuguevsky, N. Volsky, M. Kafesaki, and E.N. Economou, "Extremely high Q-factor metamaterials due to anapole excitation," Phys. Rev. B **95**(3), 035104 (2017).
[134] A. Canós Valero, D. Borovkov, A. Kalganov, A. Dudnikova, M. Sidorenko, P. Dergachev, E. Gurvitz, L. Gao, V. Bobrovs, A. Miroshnichenko, and A.S. Shalin, "On the Existence of Pure, Broadband Toroidal Sources in Electrodynamics," Laser & Photonics Reviews **18**(4), 2200740 (2024).
[135] K.J. Wo, J. Peng, M.K. Prasad, Y. Shi, J. Li, and S. Wang, "Optical forces in coupled chiral particles," Phys. Rev. A **102**(4), 043526 (2020).
[136] L. Pazin, A. Dyskin, and Y. Leviatan, "Quasi-Isotropic X-Band Inverted-F Antenna for Active RFID Tags," IEEE Antennas and Wireless Propagation Letters **8**, 27–29 (2009).
[137] C. Cho, H. Choo, and I. Park, "Broadband RFID tag antenna with quasi-isotropic radiation pattern," Electronics Letters **41**(20), 1091–1092 (2005).
[138] H.-K. Ryu, G. Jung, D.-K. Ju, S. Lim, and J.-M. Woo, "An Electrically Small Spherical UHF RFID Tag Antenna With Quasi-Isotropic Patterns for Wireless Sensor Networks," IEEE Antennas and Wireless Propagation Letters **9**, 60–62 (2010).
[139] Y. Wang, M.-C. Tang, S. Chen, L. Li, D. Li, K.-Z. Hu, and M. Li, "Design of Low-Cost, Flexible, Uniplanar, Electrically Small, Quasi-Isotropic Antenna," IEEE Antennas and Wireless Propagation Letters **18**(8), 1646–1650 (2019).
[140] Y. Wang, S. Yan, and B. Huang, "Conformal Folded Inverted-F Antenna With Quasi-Isotropic Radiation Pattern for Robust Communication in Capsule Endoscopy Applications," IEEE Transactions on Antennas and Propagation **70**(8), 6537–6550 (2022).
[141] E.M. Jung, Y. Cui, T.-H. Lin, X. He, A. Eid, J.G.D. Hester, G.D. Abowd, T.E. Starner, W.-S. Lee, and M.M. Tentzeris, "A Wideband, Quasi-Isotropic, Kilometer-Range FM Energy Harvester for Perpetual IoT," IEEE Microwave and Wireless Components Letters **30**(2), 201–204 (2020).
[142] J.-H. Kim, H. Kim, and S. Nam, "Design of a compact quasi-isotropic antenna for RF energy harvesting," in *2017 IEEE Wireless Power Transfer Conference (WPTC)*, (2017), pp. 1–3.
[143] S.I. Hussain Shah, S.M. Radha, P. Park, and I.-J. Yoon, "Recent Advancements in Quasi-Isotropic Antennas: A Review," IEEE Access **9**, 146296–146317 (2021).
[144] Y.-M. Pan, K.W. Leung, and K. Lu, "Compact Quasi-Isotropic Dielectric Resonator Antenna With Small Ground Plane," IEEE Transactions on Antennas and Propagation **62**(2), 577–585 (2014).
[145] S. Long, "A combination of linear and slot antennas for quasi-isotropic coverage," IEEE Transactions on Antennas and Propagation **23**(4), 572–576 (1975).
[146] Y. Pan, and S. Zheng, "A Compact Quasi-Isotropic Shorted Patch Antenna," IEEE Access **5**, 2771–2778 (2017).
[147] J. Ouyang, Y.M. Pan, S.Y. Zheng, and P.F. Hu, "An Electrically Small Planar Quasi-Isotropic Antenna," IEEE Antennas and Wireless Propagation Letters **17**(2), 303–306 (2018).
[148] Z. Zhang, X. Gao, W. Chen, Z. Feng, and M.F. Iskander, "Study of Conformal Switchable Antenna System on Cylindrical Surface for Isotropic Coverage," IEEE Transactions on Antennas and Propagation **59**(3), 776–783 (2011).
[149] G. Pan, Y. Li, Z. Zhang, and Z. Feng, "Isotropic Radiation From a Compact Planar Antenna Using Two Crossed Dipoles," IEEE Antennas and Wireless Propagation Letters **11**, 1338–1341

(2012).
[150] J.W. Luo, Y.M. Pan, S.Y. Zheng, and S.H. Wang, "A Planar Angled-Dipole Antenna With Quasi-Isotropic Radiation Pattern," IEEE Transactions on Antennas and Propagation **68**(7), 5646–5651 (2020).
[151] Z. Su, K. Klionovski, H. Liao, W. Li, and A. Shamim, "A Fully-Printed 3D Antenna With 92% Quasi-Isotropic and 85% CP Coverage," IEEE Transactions on Antennas and Propagation **70**(9), 7914–7922 (2022).
[152] B. Xue, and A.M.H. Wong, "Near-Field Mutual Coupling Suppression in Sub-Wavelength MIMO Janus Antenna," in *2025 IEEE International Workshop on Electromagnetics: Applications and Student Innovation Competition (iWEM)*, (2025), pp. 279–281.
[153] N. Yu, P. Genevet, M.A. Kats, F. Aieta, J.-P. Tetienne, F. Capasso, and Z. Gaburro, "Light Propagation with Phase Discontinuities: Generalized Laws of Reflection and Refraction," Science **334**(6054), 333–337 (2011).
[154] S. Sun, Q. He, S. Xiao, Q. Xu, X. Li, and L. Zhou, "Gradient-index meta-surfaces as a bridge linking propagating waves and surface waves," Nature Materials **11**, 426–31 (2012).
[155] C. Pang, Y. Wang, P. Wang, A. Yu, Y. Liu, Z. Yue, M. Hu, J. Hu, Y. Dong, and J. Qi, "Dispersion-engineered compact twisted metasurfaces enabling 3D frequency-reconfigurable holography," PhotoniX **6**(1), 33 (2025).
[156] J. Zhao, P. Zhu, Z. Wen, F. Tang, B. Zheng, R. Zhu, H. Qian, C. Qian, H. Lu, and H. Chen, "Adaptive transparent cloaking tunnel enabled by Meta-Reinforcement-Learning Metasurfaces," PhotoniX **7**(1), 2 (2026).
[157] V.A. Fedotov, P.L. Mladyonov, S.L. Prosvirnin, A.V. Rogacheva, Y. Chen, and N.I. Zheludev, "Asymmetric Propagation of Electromagnetic Waves through a Planar Chiral Structure," Phys. Rev. Lett. **97**(16), 167401 (2006).
[158] C. Menzel, C. Helgert, C. Rockstuhl, E.-B. Kley, A. Tünnermann, T. Pertsch, and F. Lederer, "Asymmetric Transmission of Linearly Polarized Light at Optical Metamaterials," Phys. Rev. Lett. **104**(25), 253902 (2010).
[159] A. Kord, D.L. Sounas, and A. Alù, "Microwave Nonreciprocity," Proceedings of the IEEE **108**(10), 1728–1758 (2020).
[160] S. Cakmakyapan, H. Caglayan, A.E. Serebryannikov, and E. Ozbay, "Experimental validation of strong directional selectivity in nonsymmetric metallic gratings with a subwavelength slit," Appl. Phys. Lett. **98**(5), 051103 (2011).
[161] J. Helszajn, "Composite-Junction Circulators Using Ferrite Disks and Dielectric Rings," IEEE Transactions on Microwave Theory and Techniques **22**(4), 400–410 (1974).
[162] X. Yang, E. Wen, and D. Sievenpiper, "All-passive microwave-diode nonreciprocal metasurface," Commun Phys **6**(1), 333 (2023).
[163] L. Zhang, X.Q. Chen, R.W. Shao, J.Y. Dai, Q. Cheng, G. Castaldi, V. Galdi, and T.J. Cui, "Breaking Reciprocity with Space-Time-Coding Digital Metasurfaces," Advanced Materials **31**(41), 1904069 (2019).
[164] Y. Li, K. Chen, J. Zhao, T. Jiang, and Y. Feng, "Emulating Faraday Rotation With Nonmagnetic Time-Modulated Metasurface," IEEE Transactions on Microwave Theory and Techniques **73**(12), 9813–9822 (2025).
[165] Y. Li, K. Duan, W. Zhao, J. Zhao, T. Jiang, K. Chen, and Y. Feng, "Tunable and Reversible Nonreciprocal Transmission with Cascaded Time-Modulated Metasurface," Laser & Photonics Reviews **20**(3), e02026 (2026).
[166] C. Huang, Y. Feng, J. Zhao, Z. Wang, and T. Jiang, "Asymmetric electromagnetic wave

transmission of linear polarization via polarization conversion through chiral metamaterial structures," Phys. Rev. B **85**(19), 195131 (2012).
[167] C. Pfeiffer, C. Zhang, V. Ray, L.J. Guo, and A. Grbic, "High Performance Bianisotropic Metasurfaces: Asymmetric Transmission of Light," Phys. Rev. Lett. **113**(2), 023902 (2014).
[168] Q. Sun, Z. Zhang, Y. Huang, X. Ma, M. Pu, Y. Guo, X. Li, and X. Luo, "Asymmetric Transmission and Wavefront Manipulation toward Dual-Frequency Meta-Holograms," ACS Photonics **6**(6), 1541–1546 (2019).
[169] B. Yao, X. Zang, Z. Li, L. Chen, J. Xie, Y. Zhu, and S. Zhuang, "Dual-layered metasurfaces for asymmetric focusing," Photon. Res., PRJ **8**(6), 830–843 (2020).
[170] K. Chen, G. Ding, G. Hu, Z. Jin, J. Zhao, Y. Feng, T. Jiang, A. Alù, and C.-W. Qiu, "Directional Janus Metasurface," Advanced Materials **32**(2), 1906352 (2020).
[171] G. Shang, H. Li, Z. Wang, S.N. Burokur, K. Zhang, J. Liu, Q. Wu, X. Ding, and X. Ding, "Transmission–Reflection-Integrated Multiplexed Janus Metasurface," ACS Appl. Electron. Mater. **3**(6), 2638–2645 (2021).
[172] K. Chen, and Y. Feng, "A review of recent progress on directional metasurfaces: concept, design, and application," J. Phys. D: Appl. Phys. **55**(38), 383001 (2022).
[173] C. Wan, C. Dai, S. Wan, Z. Li, Y. Shi, and Z. Li, "Dual-encryption freedom via a monolayer-nanotextured Janus metasurface in the broadband visible," Opt. Express, OE **29**(21), 33954–33961 (2021).
[174] E. Maguid, I. Yulevich, M. Yannai, V. Kleiner, M. L Brongersma, and E. Hasman, "Multifunctional interleaved geometric-phase dielectric metasurfaces," Light Sci Appl **6**(8), e17027–e17027 (2017).
[175] Z. Bomzon, V. Kleiner, and E. Hasman, "Pancharatnam–Berry phase in space-variant polarization-state manipulations with subwavelength gratings," Opt. Lett., OL **26**(18), 1424–1426 (2001).
[176] W. Xiao, W. Kuang, S. Huang, S. Liang, D.P. Tsai, and S. Wang, "Acoustic Pancharatnam–Berry geometric phase for structured sound manipulation," Proceedings of the National Academy of Sciences **123**(11), e2527851123 (2026).
[177] W. Xiao, S. Jia, T. Fu, and S. Wang, "Water-Wave Pancharatnam-Berry Phase Induced by 4D Spin-Orbit State Evolution," Advanced Science **13**(2), e15337 (2026).
[178] H. Pan, M.K. Chen, D.P. Tsai, and S. Wang, "Nonreciprocal Pancharatnam-Berry metasurface for unidirectional wavefront manipulations," Opt. Express, OE **32**(15), 25632–25643 (2024).
[179] F. Zhang, M. Pu, X. Li, P. Gao, X. Ma, J. Luo, H. Yu, and X. Luo, "All-Dielectric Metasurfaces for Simultaneous Giant Circular Asymmetric Transmission and Wavefront Shaping Based on Asymmetric Photonic Spin–Orbit Interactions," Advanced Functional Materials **27**(47), 1704295 (2017).
[180] J.P. Balthasar Mueller, N.A. Rubin, R.C. Devlin, B. Groever, and F. Capasso, "Metasurface Polarization Optics: Independent Phase Control of Arbitrary Orthogonal States of Polarization," Phys. Rev. Lett. **118**(11), 113901 (2017).
[181] J. Wang, K. Qu, J. Ni, W. Yang, K. Tang, S. Dong, S. Wang, J. Zhao, T. Jiang, K. Chen, and Y. Feng, "Broadband spin-unlocked achromatic meta-devices empowered by hybrid-phase cooperative dispersion engineering," PhotoniX **6**(1), 56 (2025).
[182] X. Liang, L. Deng, X. Shan, Z. Li, Z. Zhou, Z. Guan, and G. Zheng, "Asymmetric hologram with a single-size nanostructured metasurface," Opt. Express, OE **29**(13), 19964–19974 (2021).
[183] S. Dong, K. Qu, Q. Hu, S. Wang, K. Chen, and Y. Feng, "Full-Space Janus Meta-Lens for Shared-Aperture Transmission-Reflection-Independent Focusing of Electromagnetic Wave,"

Advanced Photonics Research **5**(9), 2300349 (2024).
[184] W. Yang, K. Chen, S. Dong, S. Wang, K. Qu, T. Jiang, J. Zhao, and Y. Feng, "Direction-Duplex Janus Metasurface for Full-Space Electromagnetic Wave Manipulation and Holography," ACS Appl. Mater. Interfaces **15**(22), 27380–27390 (2023).
[185] S. Dong, K. Qu, S. Wang, J. Zhao, K. Chen, and Y. Feng, "Shared-aperture full-duplex Janus meta-lens for asymmetric focusing of electromagnetic waves," Appl. Phys. Lett. **126**(2), 021703 (2025).
[186] M.-Z. Chong, C. He, P. Feng, Z.-K. Zhang, G. Geng, J. Li, M.-Y. Xia, and L. Huang, "Janus meta-imager: asymmetric image transmission and transformation enabled by diffractive neural networks," PhotoniX **6**(1), 60 (2025).
[187] B. Gholipour, J. Zhang, K.F. MacDonald, D.W. Hewak, and N.I. Zheludev, "An All-Optical, Non-volatile, Bidirectional, Phase-Change Meta-Switch," Advanced Materials **25**(22), 3050–3054 (2013).
[188] B. Chen, S. Yang, J. Chen, J. Wu, K. Chen, W. Li, Y. Tan, Z. Wang, H. Qiu, K. Fan, C. Zhang, H. Wang, Y. Feng, Y. He, B. Jin, X. Wu, J. Chen, and P. Wu, "Directional terahertz holography with thermally active Janus metasurface," Light Sci Appl **12**(1), 136 (2023).
[189] L. Bao, L.L. Zhu, X.Y. Guo, L.W. Wu, H. Zhang, P. Chang, Y.-Y. Xie, R.Y. Wu, and T.J. Cui, "Transparent and Conformal Directional Janus Metasurface with Joint Amplitude and Phase Modulations," Laser & Photonics Reviews **19**(18), 2500340 (2025).

**Table 1** Symmetry properties of the Janus dipole and evanescent reactive power under parity ($\hat{P}$) and time-reversal symmetry ($\hat{T}$).

| | $\hat{P}$ | $\hat{T}$ |
|---|---|---|
| Electric dipole | odd | even |
| Magnetic dipole | even | odd |
| **Janus dipole** | odd | even |
| Electric field | odd | even |
| Magnetic field | even | odd |
| **Reactive power** | odd | even |

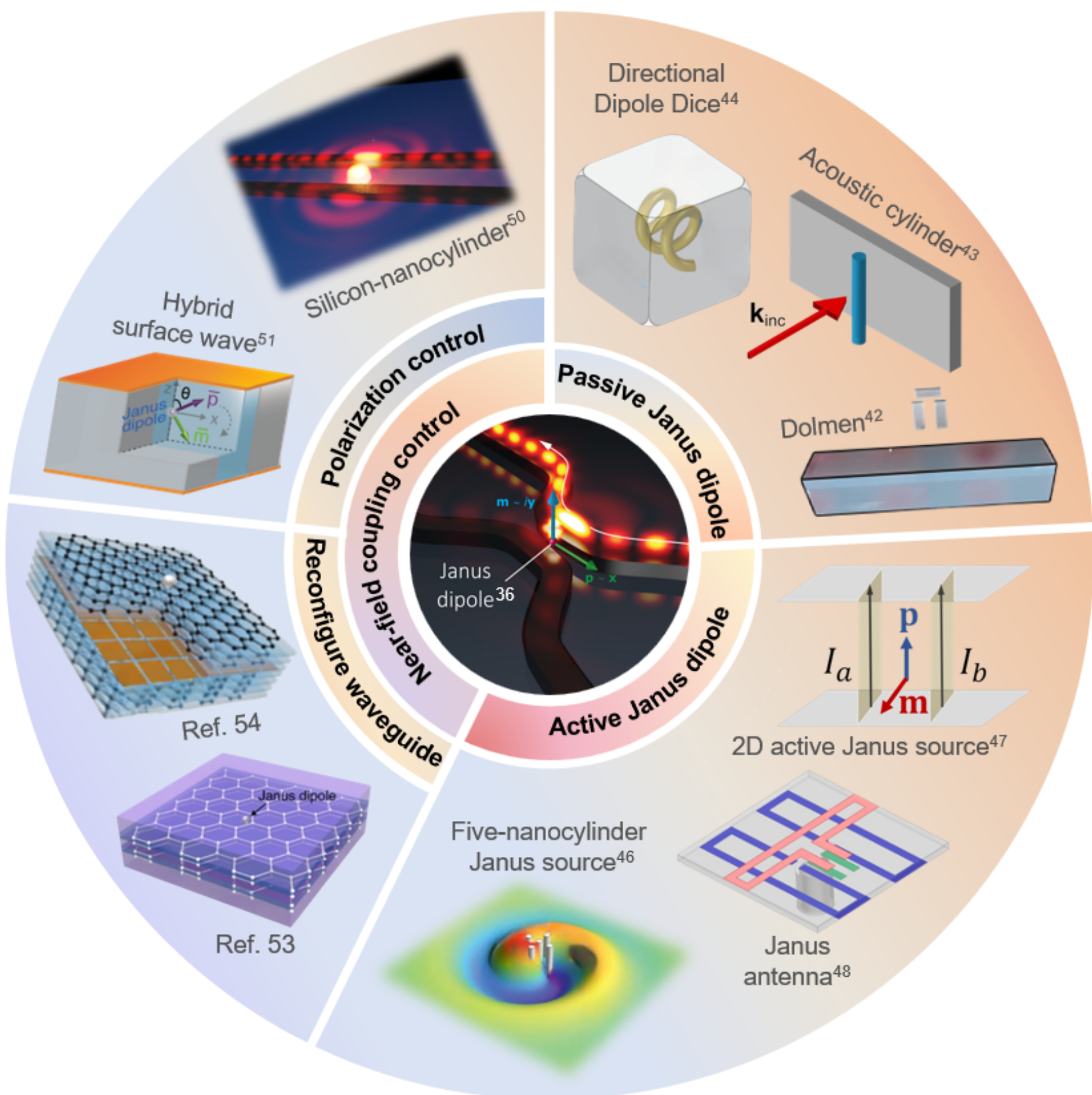

**FIG. 1.** Overview of the Janus dipole research, classified into passive Janus dipoles, active Janus dipoles, and near-field coupling control (via polarization modulation and reconfigurable waveguides).

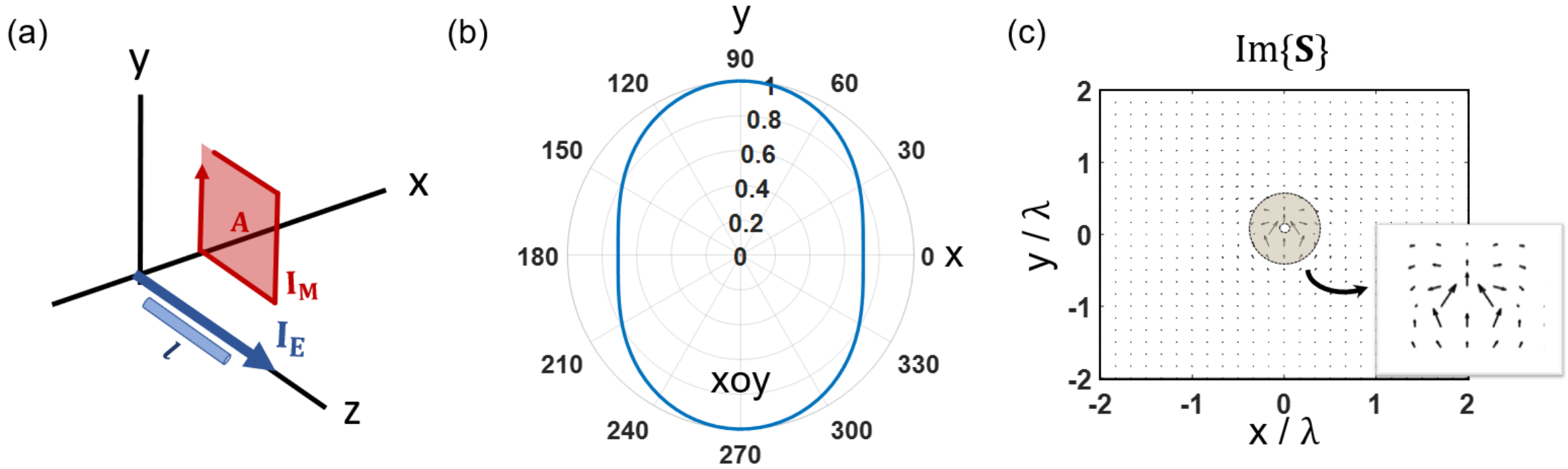

**FIG. 2.** Model of the Janus dipole and its far-field and near-field characteristics. (a) A schematic of the electric dipole of length $l$ along the z-direction and magnetic dipole formed by an electric current loop of area $A$ along the x-direction. (b) The 2D radiation pattern of the Janus source. (c) The imaginary Poynting vector $\mathrm{Im}\{\mathbf{S}\}$ for the Janus source, with a magnified inset near the source.

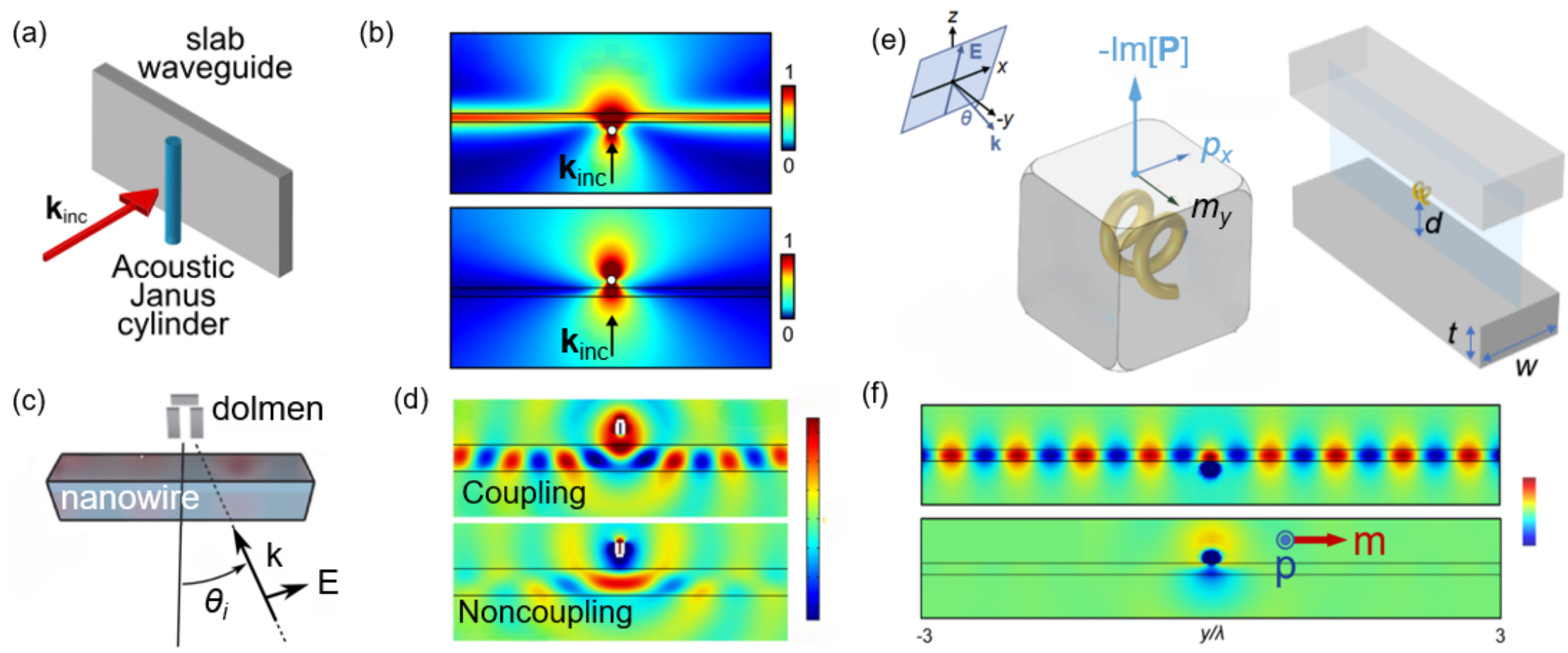

**FIG. 3.** Passive Janus dipoles. (a-b) The acoustic cylinder with slab waveguide under the incident wave and its Janus near-field coupling and non-coupling face. Reproduced with permission from Ref.[43]. Copyright (2020) Author(s), licensed under a Creative Commons Attribution (CC BY) license. (c-d) The dolmen and its Janus near-field coupling and non-coupling face. Adapted from Ref.[42]. Copyright (2020) American Physical Society. (e-f) Schematic of the Directional Dipole Dice (DDD): A single helical structure that, under tailored plane-wave illumination, simultaneously excites all three fundamental directional dipoles while enabling 3D spatial control through waveguide integration. Near-field distribution demonstrating coupling (top waveguide) and non-coupling (bottom waveguide) behavior. Adapted from Ref.[44]. Copyright (2023) Author(s), licensed under a Creative Commons Attribution (CC BY) license.

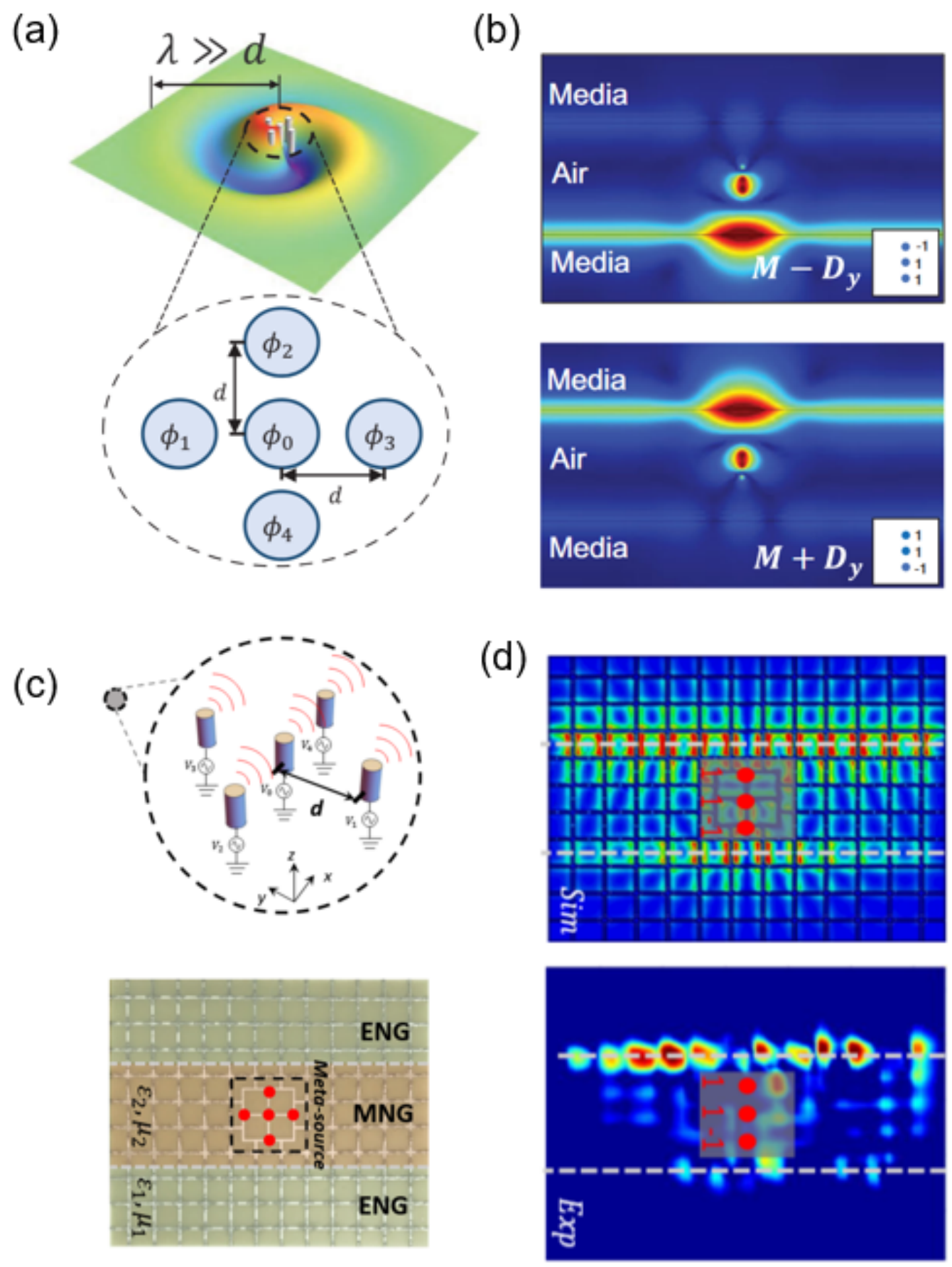

**FIG. 4.** Active Janus dipoles realized by five-source assemblies as effective Janus sources in acoustic and microwave systems. (a-b) Designed five nanocylinders assembly functioning as an effective Janus source. Interface-selective coupling behavior, showing the coupling face and non-coupling face. Reproduced with permission from Ref.[46]. Copyright (2020) Author(s), licensed under a Creative Commons Attribution (CC BY) license. (c-d) Designed a metasource comprised of five subwavelength electric sources and microwave transmission lines metamaterial comprised of epsilon-negative (ENG) and mu-negative (MNG). Simulated and measured coupling behavior, showing the coupling face and non-coupling face. Reproduced with permission from Ref.[45]. Copyright (2020) American Physical Society.

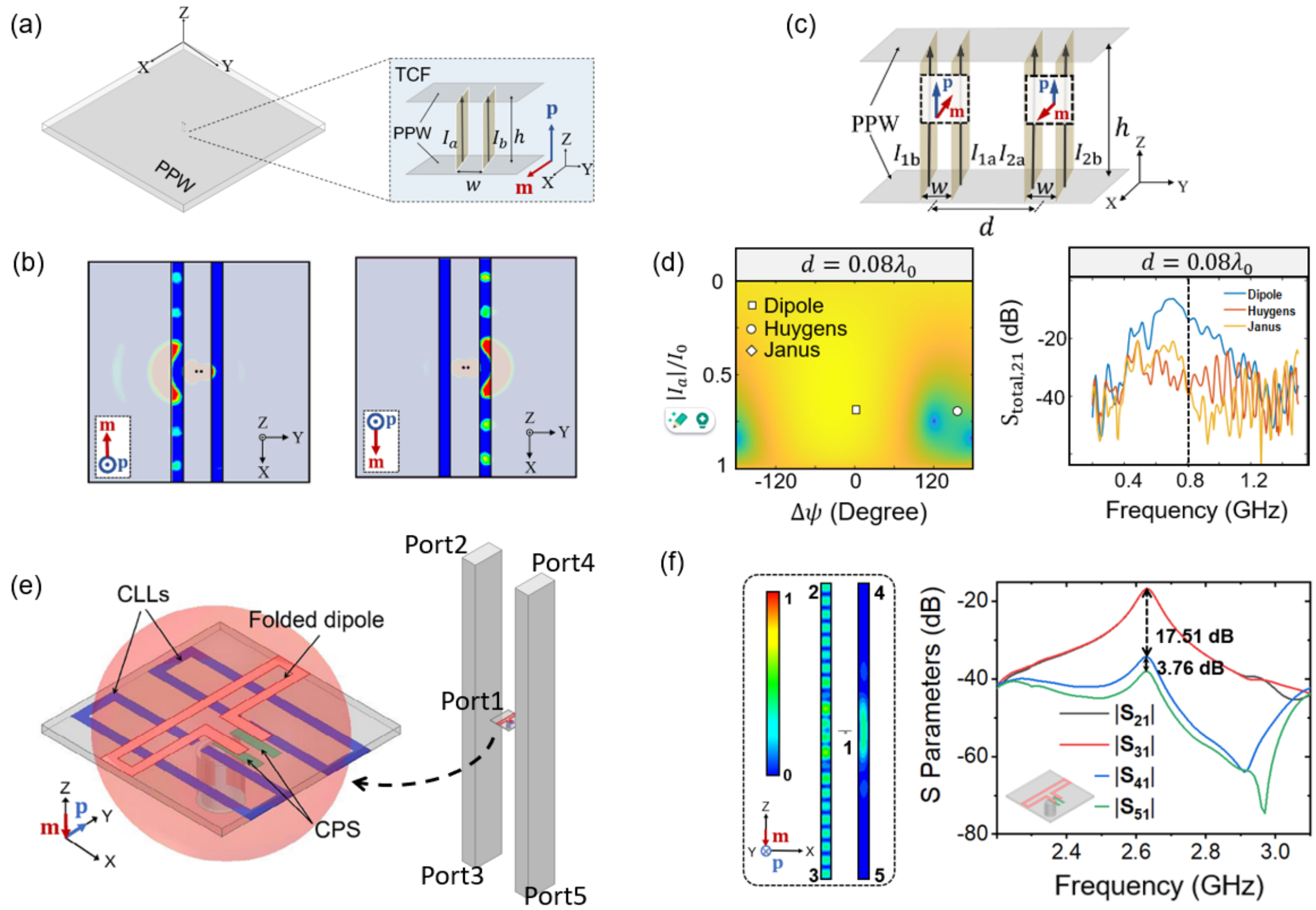

**FIG. 5.** 2D and 3D active Janus sources. (a-b) A diagram of the twin current filament (TCF) model within the parallel plate waveguide (PPW) and selective coupling of Janus source. Adapted from Ref.[47]. Copyright (2024) Author(s), licensed under a Creative Commons Attribution (CC BY) license. (c-d) A diagram of two active sources within the PPW. Simulated and measured coupling coefficients between two active sources. Adapted from Ref.[47]. Copyright (2024) Author(s), licensed under a Creative Commons Attribution (CC BY) license. (e-f) An illustration of the Janus antenna, a quasi-isotropic electrically tiny antenna made up of a driven coplanar stripline section (CPS), a folded dipole, and capacitively loaded loops (CLLs). A schematic of the Janus antenna with dielectric waveguides and the simulated near-field coupling for the optimized Janus antenna. Adapted from Ref.[48]. Copyright (2025) Author(s), licensed under a Creative Commons Attribution (CC BY) license.

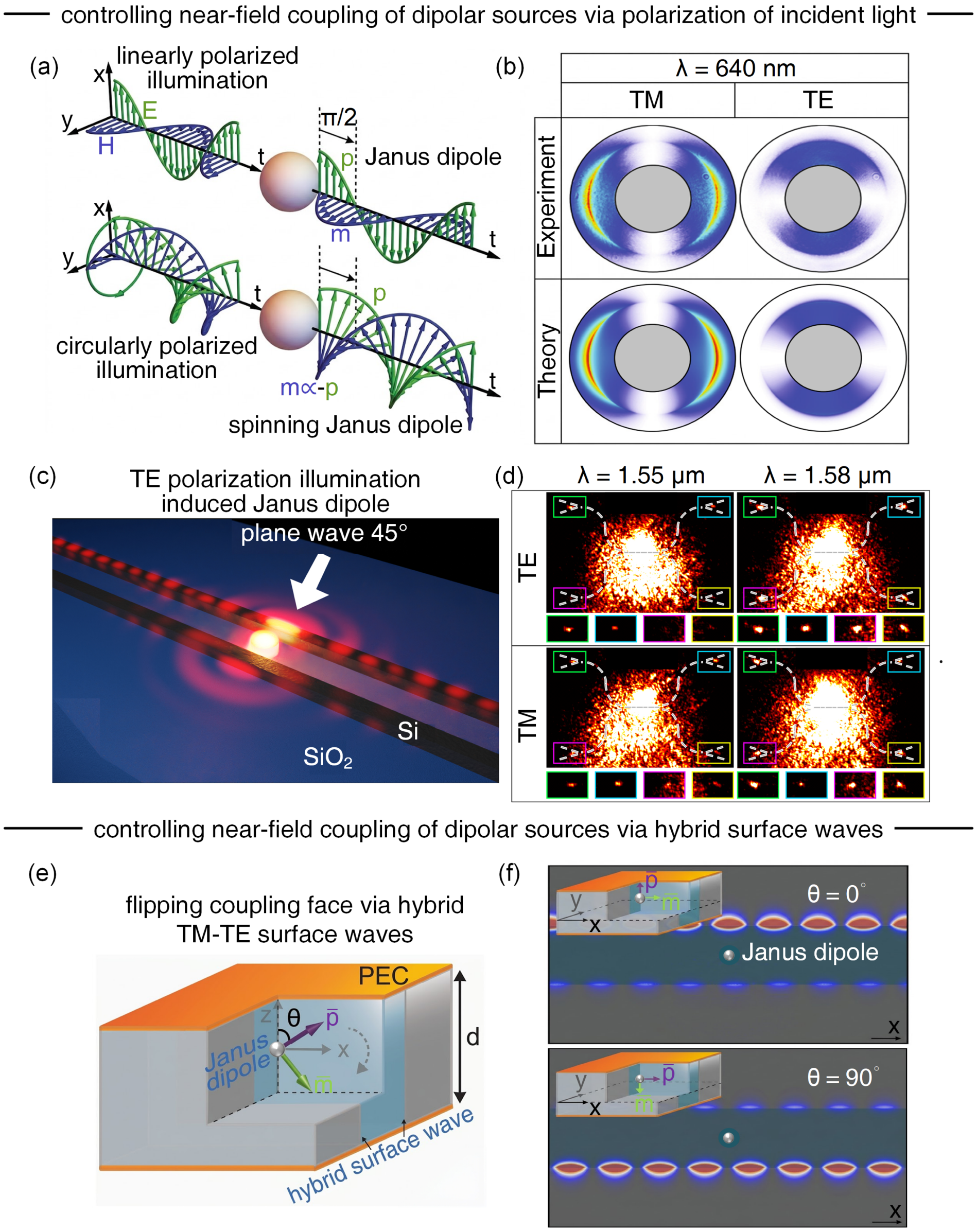

**FIG. 6.** Polarization-mediated control of the near-field coupling of Janus dipoles. (a-b) Schematic representations of a linear Janus dipole induced by linearly polarized illumination and a spinning Janus dipole induced by circularly polarized illumination. Experimental and theoretical back-focal-plane intensity distributions, showing the distinct TM and TE near-field coupling signatures. Reproduced with permission from Ref.[49]. Copyright (2019) Author(s), licensed under a Creative Commons Attribution (CC BY) license. (c-d) Schematic of an integrated silicon-nanocylinder Janus source positioned between two silicon waveguides on a $SiO_2$ substrate and excited by a TE plane wave incident at 45°. Experimentally measured outcoupling patterns under TE and TM illumination, demonstrating polarization- and wavelength-selective routing. Reproduced with permission from Ref.[50]. Copyright (2022) Author(s), licensed under a Creative Commons Attribution (CC BY) license. (e-f) Configuration for reversing the coupling face of a Janus dipole through hybrid TM-TE surface waves supported inside a parallel-plate waveguide. Simulated field distributions for dipole rotation angles $\theta = 0°$ and $90°$, showing the reversal of the coupling and non-coupling faces without changing the surface-wave polarization. Reproduced with permission from Ref.[51]. Copyright (2024) Wiley-VCH.

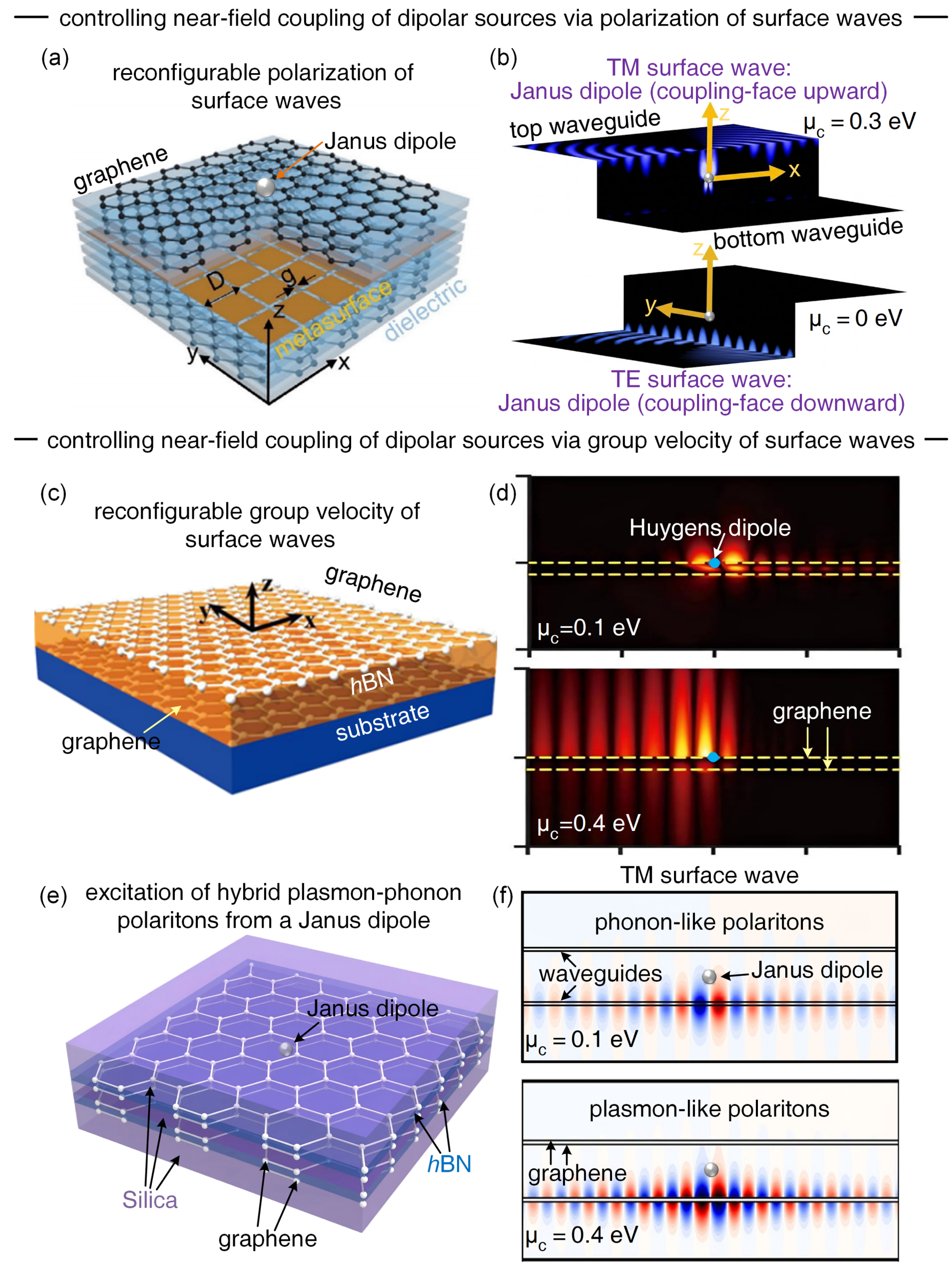

**FIG. 7.** Controlling near-field coupling of Janus dipole via reconfigurable waveguides. (a-b) Schematic of a graphene-metasurface waveguide supporting electrically reconfigurable surface-wave polarization. Reversal of the coupling face of a Janus dipole. Reproduced with permission from Ref.[54]. Copyright (2021) Wiley-VCH. (c-d) Graphene-*h*BN heterostructure supporting hybrid plasmon-phonon polaritons whose group velocity can be reconfigured by tuning the graphene chemical potential. Field distributions excited by a Huygens dipole, illustrating the change in energy-flow direction associated with the reversal of the polariton group velocity. Reproduced with permission from Ref.[52]. Copyright (2018) Wiley-VCH. (e-f) Five-layer graphene-*h*BN waveguide excited by a Janus dipole. Corresponding phonon-like and plasmon-like polariton field distributions. Adapted from Ref.[53]. Copyright (2021) The Electromagnetics Academy.

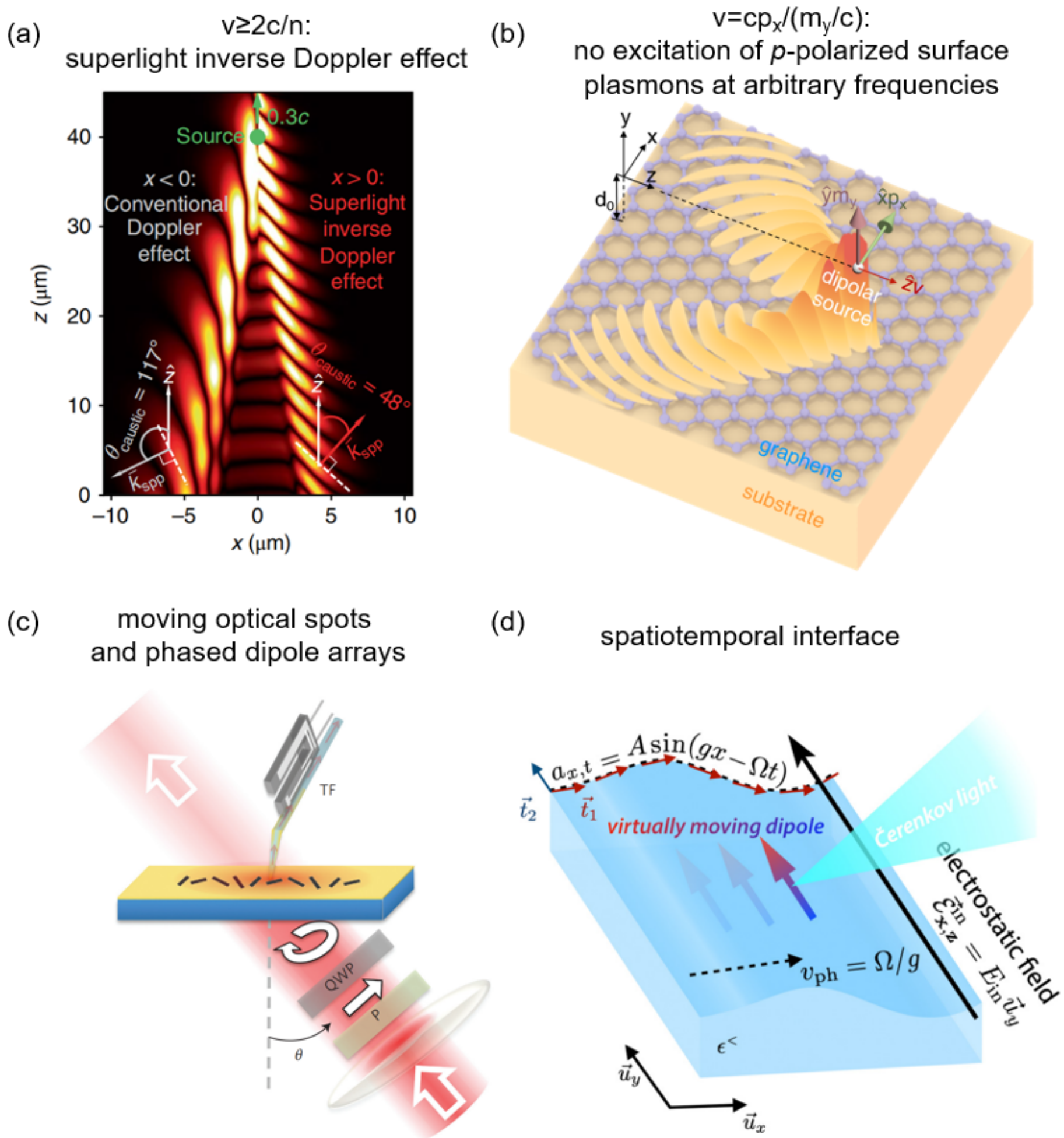

**FIG. 8.** Critical velocities and realization of moving dipolar sources. (a) Spatial separation of conventional Doppler effect and superlight inverse Doppler effect of moving circularly polarized dipoles. Reproduced with permission from Ref.[115]. Copyright (2018) Springer Nature. (b) Critical polarization suppression in the near-field interference of moving Huygens-like dipoles. Reproduced with permission from Ref.[117]. Copyright (2025) The Optical Society. (c) Realization of moving dipolar sources by moving optical spots and phased dipole arrays. Reproduced with permission from Ref.[126]. Copyright (2015) Springer Nature. (d) Realization of moving dipolar sources by spatiotemporal interfaces. Reproduced with permission from Ref.[124]. Copyright (2022) Author(s), licensed under a Creative Commons Attribution (CC BY) license.

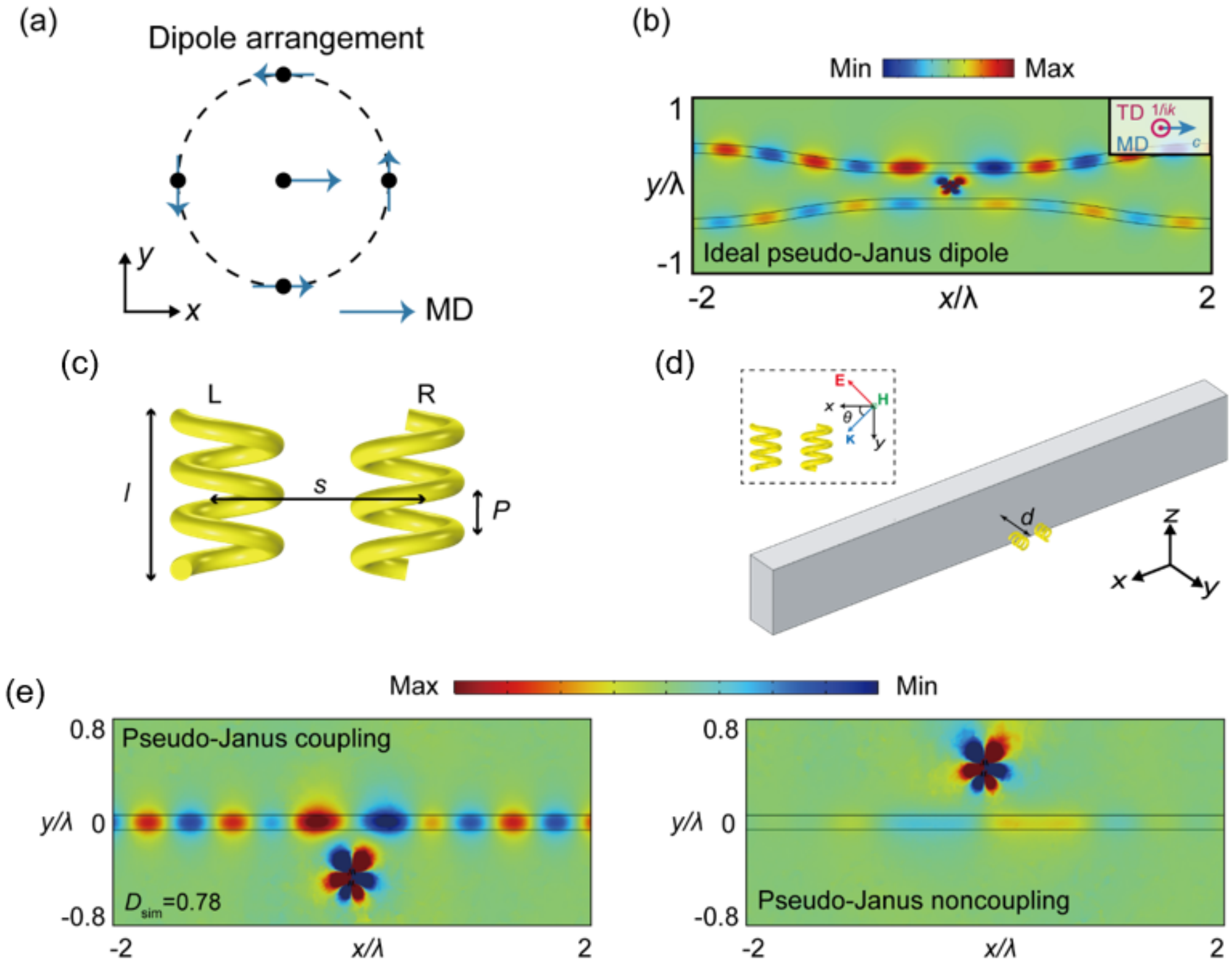

**FIG. 9.** Toroidal (higher-order) Janus dipoles. (a-b) Arrangement of active magnetic dipoles to realize a pseudo-Janus dipole. Magnetic field distribution induced by the pseudo-Janus dipole. The dipole is sandwiched by two identical bent waveguides. Reproduced with permission from Ref.[40]. Copyright (2024) American Physical Society. (c-e) A pair of helices for realizing a passive pseudo-Janus dipole. The two helices have opposite handedness, denoted by L and R. Waveguide system for demonstrating the directional coupling of the passive pseudo-Janus dipole. The helices are located at a distance $d$ from the surface of a silicon waveguide. Magnetic field distribution induced by the coupling (left) and noncoupling sides of the pseudo-Janus dipole. Reproduced with permission from Ref.[41]. Copyright (2025) Author(s), licensed under a Creative Commons Attribution (CC BY) license.

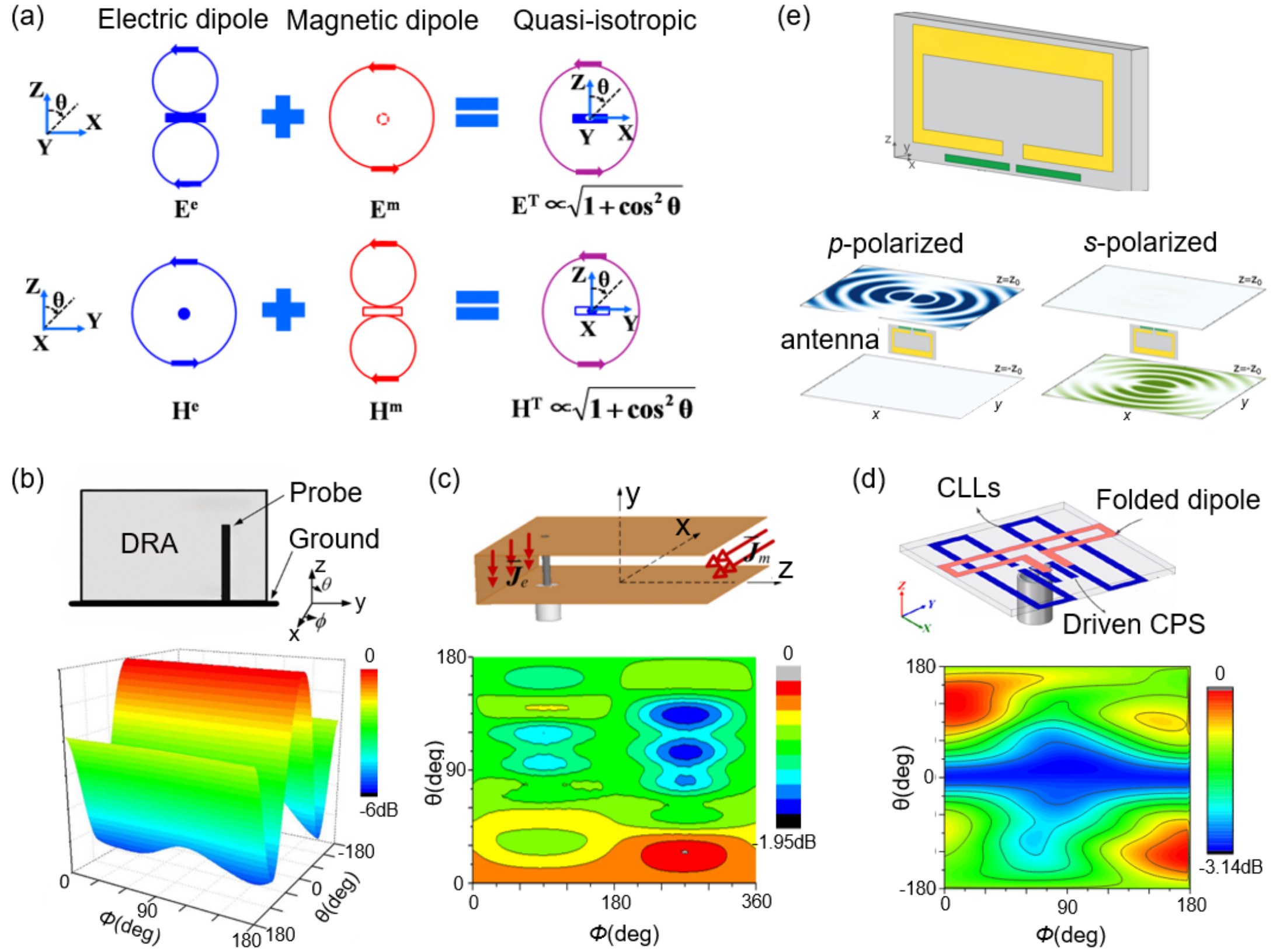

**FIG. 10.** Quasi-isotropic antennas and Janus dipole antenna. (a) Orthogonal design concepts of the quasi-isotropic radiation pattern. Adapted from Ref.[139]. Copyright (2019) IEEE. (b) Quasi-isotropic dielectric resonator antenna. Adapted from Ref.[144]. Copyright (2014) IEEE. (c) Quasi-isotropic shorted patch antenna. Adapted from Ref.[146]. Copyright (2017) IEEE. and (d) Electrically small quasi-isotropic antenna. Adapted from Ref.[147]. Copyright (2018) IEEE. (e) Janus dipole antenna structure and the p-polarized guided waves are excited almost only in the upper waveguide, while the s-polarized guided waves are excited almost only in the lower waveguide. Adapted from Ref.[55]. Copyright (2025) American Physical Society.

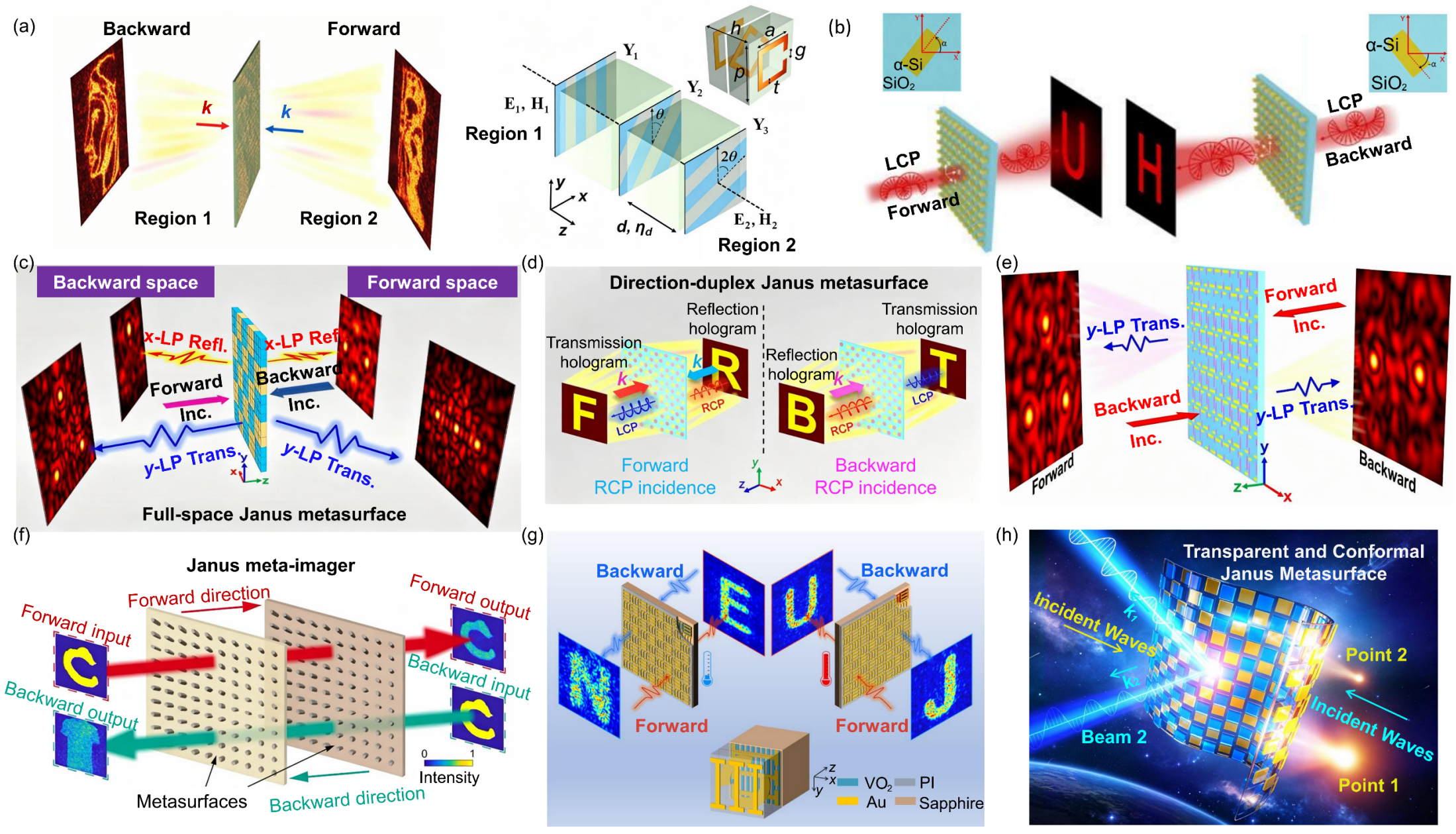

**FIG. 11.** Advanced Janus metasurfaces for multifunctional wavefront tailoring. (a) Directional Janus metasurface for linearly polarized waves using cascaded twisted meta-atoms. Adapted from Ref.[170]. Copyright (2020) Wiley-VCH. (b) Asymmetric hologram based on single-size silicon nanobricks utilizing geometric phase. Adapted from Ref.[182]. Copyright (2021) Author(s), licensed under a Creative Commons Attribution (CC BY) license. (c) Full-space Janus meta-lens for independent focusing. Adapted from Ref.[183]. Copyright (2024) Author(s), licensed under a Creative Commons Attribution (CC BY) license. (d) Direction-duplex Janus metasurface with interleaved meta-atoms. Adapted from Ref.[184]. Copyright (2023) American Chemical Society. (e) Shared-aperture full-duplex Janus meta-lens using receiver–transmitter meta-atoms. Adapted from Ref.[185]. Copyright (2025) AIP Publishing. (f) Janus meta-imager enabled by diffractive deep neural networks. Adapted from Ref.[186]. Copyright (2025) Author(s), licensed under a Creative Commons Attribution (CC BY) license. (g) Thermally active Janus metasurface using $VO_2$. Adapted from Ref.[188]. Copyright (2023) Author(s), licensed under a Creative Commons Attribution (CC BY) license. (h) Transparent and conformal Janus metasurface on flexible PET-ITO films. Adapted from Ref.[189]. Copyright (2025) Wiley-VCH.